\documentclass[useAMS,usenatbib]{mnras}
\usepackage{rotating, graphicx}
\usepackage{amsmath}
\usepackage{amssymb} 

\newcommand{\Msol}{\,{\rm M}_{\odot}}
\newcommand{\Mpc}{\,{\rm Mpc}}
\newcommand{\Gyr}{\,{\rm Gyr}}
\newcommand{\kms}{\,{\rm km}\,{\rm s}^{-1}}

\newcommand{\K}{\,{\rm K}}
\newcommand{\cm}{\,{\rm cm}}
\newcommand{\pkpc}{\,{\rm pkpc}}
\newcommand{\pc}{\,{\rm pc}}
\newcommand{\erg}{\,{\rm erg}}
\newcommand{\keVcm}{\,{\rm keV}\,{\rm cm}^{2}}
\newcommand{\percm}{\,{\rm cm}^{-3}}

\newcommand{\Mbh}{M_{\rm bh}}
\newcommand{\Mhalo}{M_{\rm halo}}
\newcommand{\oneplusz}{\Delta_z}   % {1+z}

\newcommand{\rgb}{}

\title[The dark nemesis of galaxy formation]{The dark nemesis of galaxy formation: why hot haloes trigger black hole growth and bring star formation to an end}
\author[R. G. Bower et al]{Richard G. Bower$^1$\thanks{E-mail: r.g.bower@dur.ac.uk}, 
Joop Schaye$^2$, Carlos S. Frenk$^1$, Tom Theuns$^1$, \and 
Matthieu Schaller$^1$,   Robert A. Crain$^3$, Stuart McAlpine$^1$.\\
 $^1$ Institute for Computational Cosmology, Department of Physics, University of Durham, South Road, Durham, DH1 3LE, UK\\
 $^2$ Leiden Observatory, Leiden University, PO Box 9513, 2300 RA Leiden, The Netherlands\\
 $^3$ Astrophysics Research Institute, Liverpool John Moores University, 146 Brownlow Hill, Liverpool, L3 5RF, UK}

\date{Accepted XXX. Received YYY; in original form ZZZ}

\pubyear{2016}

\begin{document}
\label{firstpage}
\pagerange{\pageref{firstpage}--\pageref{lastpage}}
\maketitle

\begin{abstract}
Galaxies fall into two clearly distinct types: `blue-sequence' galaxies that are rapidly forming young stars, and `red-sequence' galaxies in which star formation has almost completely ceased. Most galaxies more massive than $3\times10^{10}\Msol$ follow the red-sequence while less massive central galaxies lie on the blue sequence. We show that these sequences are created by a competition between star formation-driven outflows and gas accretion on to the supermassive black hole at the galaxy's centre. We develop a simple analytic model for this interaction. In galaxies less massive than $3\times10^{10}\Msol$ young stars and supernovae drive a high entropy outflow
which is more buoyant than any tenuous corona. The outflow balances the rate of gas inflow, preventing high gas densities building up in the central regions. More massive galaxies, however, are surrounded by an increasingly hot corona. Above a halo mass of $\sim 10^{12}\Msol$, the outflow ceases to be buoyant and star formation is unable to prevent the build up of gas in the central regions. This triggers a strongly non-linear response from the black hole. Its accretion rate rises rapidly, heating the galaxy's corona, disrupting the incoming supply of cool gas and starving the galaxy of the fuel for star formation. The host galaxy makes a transition to the red sequence, and further growth predominantly occurs through galaxy mergers. We show that the analytic model provides a good description of galaxy evolution in the EAGLE hydrodynamic simulations. So long as star formation-driven outflows are present, the transition mass scale is almost independent of  subgrid parameter choice.
\end{abstract}

\begin{keywords}
black hole physics, galaxies: formation, galaxies: active, 
methods: hydrodynamical simulations, quasars: general.
\end{keywords}

\section{Introduction}

Galaxies fall into two clearly distinct types: active, `blue-sequence' galaxies that are rapidly forming young stars, and passive `red-sequence' galaxies in which star formation has almost completely ceased. The two sequences are clearly seen when galaxy colours or star formation rates are plotted as a function of galaxy mass (eg., \citet{kauffmann2003, baldry2006}).
Low-mass galaxies generally follow the `blue-sequence' with a tight, almost linear, relationship between star formation rate and stellar mass \citep[eg.,][]{brinchmann2004}, while massive galaxies follow the `red-sequence' with almost undetectable levels of star formation \citep[eg.,][]{bower1992}. %bell2004}.
At the present day, the transition between the two types occurs at a stellar mass scale of $3\times10^{10}\Msol$.  Galaxies less massive than the transition-scale grow through star formation, doubling their stellar mass on a timescale comparable to the age of the Universe; above the transition mass, galaxy growth slows and is driven primarily by galaxy mergers  \citep[eg.,][]{delucia2006,parry2009,qu2016,rodriguez2016}. 
The existence of the transition mass is closely related to the form of the galaxy stellar mass 
function, creating the exponential break at high masses \citep[eg.,][]{benson2003, peng2010}. The transition mass is sometimes referred to as the `quenching' mass scale. In 
this paper we will focus on the properties of central galaxies (we will not consider the environmental effects that may suppress star formation in lower mass satellite galaxies, see \citet{trayford2016}) and show that the transition mass scale arises from a competition between star formation driven outflows and black hole accretion.

\begin{figure*}
\centering
\includegraphics[width=\linewidth]{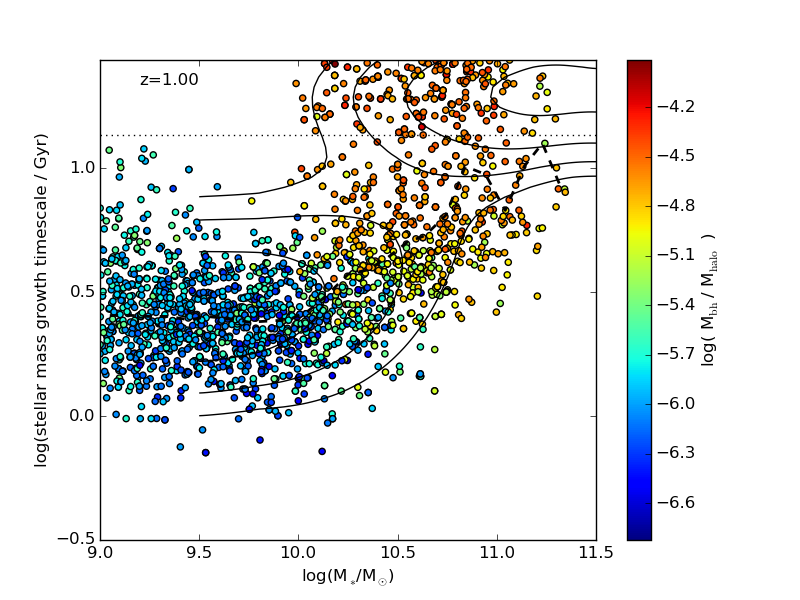}
\caption{
The formation timescales of galaxies, $M_*/\dot{M}_*$, as a function of stellar mass. Contours show observational data for galaxies at $z=1$ \citep{ilbert2015}, using the revised absolute star formation rate calibration of \citet{chang2015}.  The separation of galaxies into blue (rapidly star forming) and red (passive) sequences is clearly seen. Most low-mass galaxies follow the star forming blue galaxy sequence, doubling their stellar mass every 3 billion years, but more massive galaxies have much longer star formation growth timescales. The horizontal dotted line shows the present day age of the Universe; galaxies with longer star formation timescales are randomly placed above the line. The transition between the sequences occurs at a stellar mass of round $3\times10^{10}\Msol$, similar to the transition mass scale observed in the present-day Universe. We supplement the observational data with model galaxies from the reference EAGLE cosmological simulation (filled circles). The simulated galaxies follow the observed data closely. The points are coloured by the mass of the black hole relative to that of the host halo. Around the transition mass scale, there is considerable scatter in the relative mass of the black hole and in the star formation growth timescale. However, at a fixed galaxy mass, systems with a higher black hole mass tend to have substantially longer growth timescales, implying the existence of an intimate connection between black hole mass and galaxy star formation activity.
}
\label{fig:sfgrowth_mstar}
\end{figure*}

Most of the stars in the Universe today were, however, formed when the Universe was less than half its present age.
Recently, deep redshift surveys have been able to convincingly demonstrate the existence of a transition mass at higher redshifts  \citep[eg.,][]{peng2010,ilbert2015,darvish2016}. It is useful to illustrate the balance of galaxies on the two sequences by revisiting the analysis
of \citet{kauffmann2003}. This is illustrated in Fig.~\ref{fig:sfgrowth_mstar},  where contours show the star formation rate growth timescale of galaxies at $z=1$ using observational data from the COSMOS high-redshift galaxy survey \citep{ilbert2015}. 
So that passive galaxies appear in the figure, we assign galaxies without detectable star formation a growth timescale of $20\Gyr$ with a scatter of 0.2 dex. The relative contributions of blue-and red-sequence galaxies are then computed using the luminosity functions presented by \citet{ilbert2013}. The figure clearly shows that the division into sequences seen in present-day galaxies was already established 8 billion years ago and that the transition mass scale of $3\times10^{10}\Msol$ has changed little over the intervening time. 

%% a simple model
A simple way to begin to understand galaxy formation is to view galaxies as equilibrium systems in
which the star formation rate must balance the gas inflow rate, either by converting the inflowing gas into stars or, more importantly, by driving an outflow \citep[eg.,][]{white_frenk1991,finlator2008,schaye2010,bouche2010,dave2012}. Such a model broadly explains many aspects of galaxy evolution, such as the almost linear correlation between stellar mass and star formation rate, and the rate of evolution of this sequence. In order to explain the flat faint-end slope of the 
galaxy mass function, such models require that the mass loading of the outflow depends strongly on galaxy mass. Low-mass galaxies lose most of their
mass in the outflow (and form few stars) while galaxies around the transition mass scale consume much of the inflowing mass in star formation.

%% black holes in the quasi-equilibrium model%
Such models do not generally consider the role of the nuclear supermassive black hole, however. Yet the energy liberated when one solar mass of gas is accreted by a black hole is 10,000 times the supernova energy released by forming the same mass of stars. 
Observational measurements of energetic outflows and radio jets from galaxy nuclei \citep[eg.,][]{harrison2012, maiolino2012, fabian2003} indeed suggest that black holes play an important role in galaxy formation, most likely by heating the surrounding gas corona, offsetting cooling losses and disrupting the gas inflow \citep[eg.,][]{binney_tabor1995, silk_rees1998, dubois2013}.  Such observations have motivated the inclusion of black hole feedback in cosmological galaxy formation models \citep[eg.,][]{dimatteo2005,croton2006,bower2006,sijacki2007,booth2009,dubois2013,sijacki2015,dubois2016}. 

The success of these models results from including two different modes of black hole feedback or accretion depending on halo mass or Eddington accretion ratio. In the most extreme implementation, the effect of black hole feedback is implemented by switching cooling off in massive haloes \citep{gabor2015}. There are three possible arguments that may be advanced to support a halo mass or accretion rate dependence: (i) (in the absence of disk instabilities) black holes are only able to accrete efficiently if the surrounding gas is hot and pressure supported \citep{croton2006}; (ii) black hole
feedback is only effective if the black hole is surrounded by a hot halo that is able to capture the energy of the black hole jet \citep{bower2006}; (iii) only black holes accreting well below the Eddington limit (perhaps in the ADAF regime) are able to generate mechanical outflows \citep{meier1999,nemmen2007}. 
In the latter case, a halo mass dependence of the feedback efficiency emerges because of the strong dependence of black hole mass on halo mass.
In each case, however, the models provide only a qualitative explanation of the transition mass scale. The first case assumes
angular momentum prevents accretion of cold gas; the second argument does not explain why black holes do not undergo run-away
growth in low-mass haloes; the third links galaxy properties to the highly uncertain AU-scale physics of black hole accretion disks. 

In contrast to models that include an explicit mass or accretion rate dependence, the MassiveBlackII \citep{khandai2015} and EAGLE \citep{schaye2015} simulations  adopt a simpler description in which AGN feedback power is a fixed fraction of the rest mass accretion rate. Despite this simple proportionality, the EAGLE simulations reproduce 
the observed galaxy transition mass scale well. For example, the star formation growth time scales of EAGLE galaxies, shown by the coloured points in Fig~\ref{fig:sfgrowth_mstar}, match the observation data well (for further discussion, see \citet{furlong2015}). Although a transition mass is not so clearly evident in the MassiveBlackII simulations, the results are broadly consistent with a variation of the EAGLE simulation in which star formation driven outflow are weak (see \S\ref{sec:parameter_variations}). 

The success of the simple black hole feedback scheme in EAGLE suggests that the transition mass scale emerges from the interaction between star formation feedback and black hole fueling. This has motivated us to explore a scenario in which star formation-driven outflows themselves regulate the density of gas reaching the black hole. This has previously been considered in the context of high resolution simulations of individual high redshift galaxies \citep[eg.,][]{dubois2015, habouzit2016}.
In low-mass galaxies, such outflows are efficient, making it difficult for high gas densities to build up in the central regions of the galaxy. The critical and novel component of our model is the recognition that outflows from more massive galaxies are not buoyant. 
As the mass of the galaxy's dark matter halo increases, a hot corona begins to form \citep[eg.,][]{white_frenk1991, birnboim2003}
and the outflow becomes trapped. Star formation can then no longer prevent the gas density increasing in the central regions of the galaxy, triggering a strongly non-linear response from the black hole, heating the corona and disrupting the inflow of fresh fuel for star formation. In this picture, the falling effectiveness of feedback from star formation leads to an increase in accretion on to the black hole. Black hole feedback will take over just as star formation driven feedback fails.

The structure of the paper is as follows.
In \S\ref{sec:analytic_model}, we develop a simple analytic model that captures the key physics of the problem. We begin by exploring the strongly non-linear growth rates of black holes accreting from a constant density medium. We show that in the Bondi accretion regime black holes initially grow slowly and then abruptly switch to a rapid accretion phase. In the absence of feedback, the black hole grows to infinite mass in a finite time. We go on to consider how the density in the central parts of the galaxy will evolve as the halo mass grows, and develop a model in which the gas density is determined by the buoyancy of the star formation driven outflow. We show that a critical mass scale emerges: below this scale galaxies balance gas inflow with star formation driven outflows, but above this mass the galaxy and its gas corona are regulated by black hole accretion and star formation is strongly suppressed.
We use this simple model to explore the growth of black holes in a cosmological context.
In \S\ref{sec:validation}, we validate the analytic model by comparing it to galaxies forming in the EAGLE hydrodynamic simulation suite. We show that the approximations of the analytic model are well supported, and explore the effects of varying the sub-grid parameters of the simulations. We show that the galaxy transition mass scale is robust to the choice of subgrid parameters, but that it disappears entirely if star formation
driven outflow are absent. We present a summary of the results and a discussion of the model's wider implications in \S\ref{sec:discussion}.

\section{An Analytic Model For the Growth of Black Holes}
\label{sec:analytic_model}

\begin{figure*}
\centering
\includegraphics[width=\linewidth]{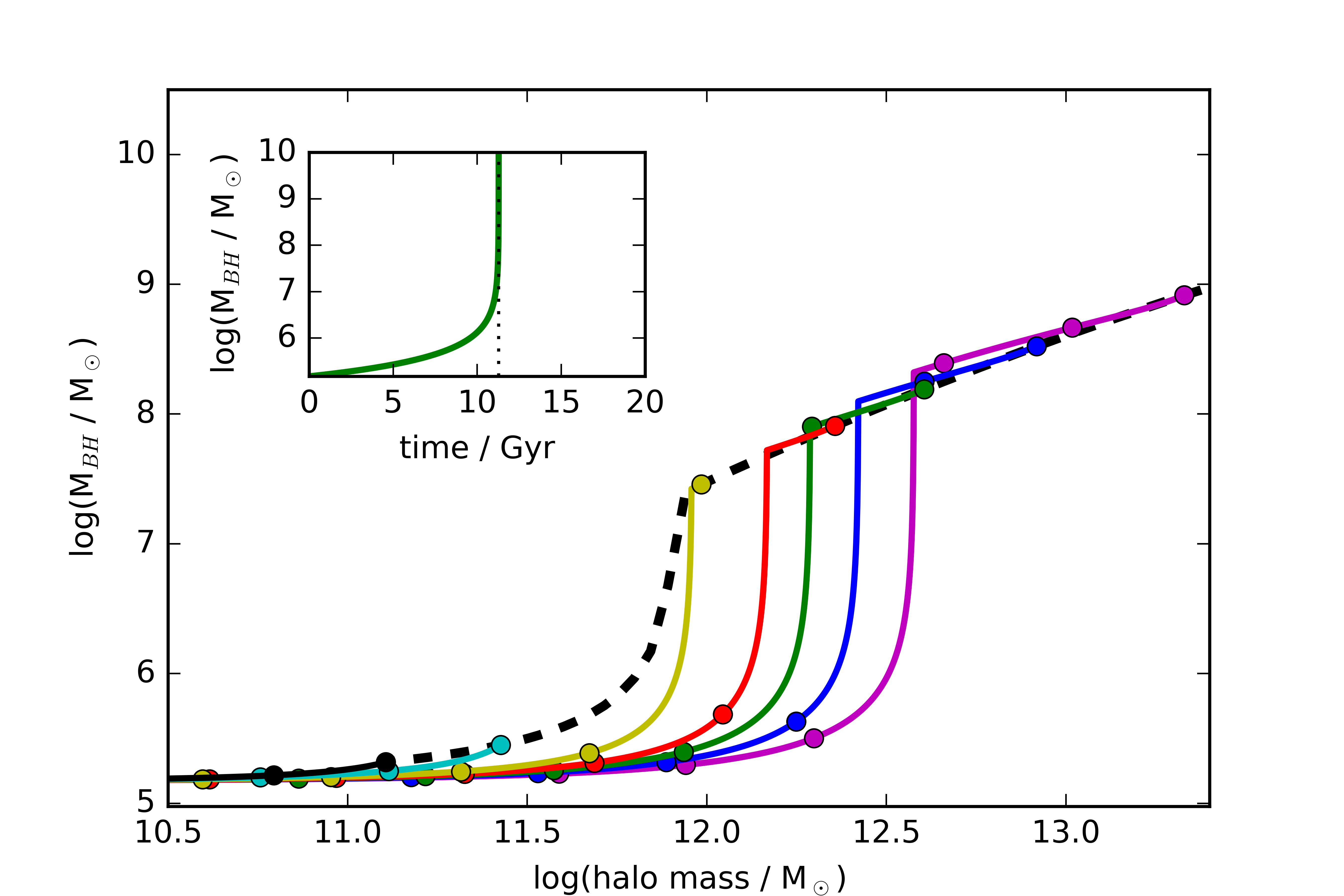}
\caption{
The non-linear growth of black holes. {\it Inset panel:} The evolution of the mass of a black hole as a function of time, assuming a constant density gas environment  (Eq.~\ref{eq:bh_mdot}). The vertical dotted line denotes the time, $t_{\infty}$, at which the black hole mass would become infinite (Eq.~\ref{eq:t_infinty}). The non-linearity of the accretion rate means that the black hole spends most of its time in the low-accretion rate phase and then suddenly switches to a rapid growth phase. 
In the {\it main panel} we show the evolution of black hole mass as a function of dark matter halo mass, assuming that the halo growth follows Eq.~\ref{eq:halogrowthrate} and that the density of gas surrounding the black hole is given by Eq.~\ref{eq:rhoBH}. We assume that the black hole grows until its energy output exceeds the halo binding energy (see \S\ref{sec:bh_mass_limit}). Coloured lines show the growth of black holes that are created when the Universe is between 0.4 and 2 Gyr old (purple to cyan, respectively) as seeds with mass $1.5\times10^5\Msol$ in haloes of mass $10^{10}\Msol$. 
By connecting the final masses of black holes created at different times, we obtain the predicted relation between black hole mass and dark matter halo mass at the present-day. This is shown by the black dashed line. 
\label{fig:bh_growth_model}
}
\end{figure*}

\subsection{Black hole growth in a constant density medium}
\label{sec:const_density}
We begin by considering the growth of a black hole in a constant density medium. Black holes accreting at their theoretical maximum rate (the Eddington rate)
grow exponentially with a timescale,  $t_{\rm salp} = 4.5\times10^{7} \,{\rm yr}$ (since $\dot{M}_{\rm bh}\propto\Mbh$) that is much shorter than the present-age of the Universe \citep{salpeter1964} .
In practice, however, the accretion rates of black holes usually lie below this rate because the gas density is low.
If the gas surrounding the black hole has density $\rho_{\rm bh}$ and effective sound speed (including turbulent pressure of the interstellar medium), $c_s$, the black hole accretion disk is fed at a rate \citep{bondi1952}:
\begin{equation} 
\dot{M}_{\rm bh} =  4\pi {\rm G}^2 f_{\rm sup} \frac{\Mbh^2\rho_{\rm bh}}{c_s^{3}} . 
\label{eq:bh_mdot}
\end{equation}
(where $\rm G$ is Newton's gravitational constant).
The relevant scale on which the density is measured is given by the Bondi radius, corresponding to around 500 pc for a black hole of
mass $10^7 \Msol$ accreting from the diffuse interstellar medium (assuming an effective sound speed of $10\kms$). We have inserted a factor, $f_{\rm sup}$, to account for the suppression of accretion by the gas motion relative
to the black hole and angular momentum \citep{bondi_hoyle1944}; on average, $f_{\rm sup} \sim 0.1$ in the EAGLE simulations irrespective of the system mass \citep{rosas2015}. We will later validate our analytic model by comparing to the EAGLE simulations that assume star forming gas follows an effective equation of state $P_{\rm eff} \propto \rho^{4/3}$ when averaged over the kpc-scale areas of the galaxy.
In this case, $\dot{M}_{\rm bh}  \propto \Mbh^2 \rho_{\rm bh}^{1/2}$. Assuming, instead, an isothermal equation of state (ie., $c_s$ constant) would result in a stronger density dependence, $\dot{M}_{\rm bh} \propto \Mbh^2 \rho_{\rm bh}$, which would strengthen our conclusions.

In the Bondi regime, black hole accretion is highly non-linear: as the black hole mass doubles, the rate of accretion increases four-fold, so that the growth timescale decreases rapidly as the black hole grows. If the density of the external medium remains constant, the black hole will grow to an infinite mass in a finite time given by:
\begin{equation}
t_{\infty} = \frac{1} {\kappa \rho_{\rm bh}^{1/2}} 
\label{eq:t_infinty}
\end{equation}
where $\kappa$ depends on the effective equation of state of the interstellar medium (ISM), the relative motion of the surrounding gas, and the initial black hole seed mass.
%\footnote{For the EAGLE reference model, $\kappa = f_{\rm mot} 4\pi G M_{\rm seed} \rho_0^{1/2} \csnorm^{-3}$ where  $\rho_0 = 3.3\times10^{15} \Msol\Mpc^{-3}$ (corresponding to a hydrogen number density of 0.1 Hydrogen atoms per cm$^{3}$) is the density threshold above which the equation of state is applied and $\csnorm= 9.50 \kms$ is the corresponding sound speed normalisation.}.  
This behaviour is illustrated in the inset to Fig.~\ref{fig:bh_growth_model}. In practice, of course, the black hole does not grow indefinitely, since it will eventually exhaust or expel the surrounding gas. % \citep{booth_schaye2010}.
This sets an upper limit to the rapid growth of the black hole, and we assume that further black hole growth is limited by the binding energy within the cooling radius of the dark matter halo.  We estimate this limiting mass in \S\ref{sec:bh_mass_limit}.  Because the coefficient $\kappa$ depends on the initial seed
mass, we might expect the growth of black holes in the EAGLE simulations to be very sensitive to this parameter.  We find, however, that a reduction in the initial black hole mass is compensated by an increase in the density around the black hole (\S\ref{sec:parameter_variations}). As a result, the halo
mass at which black holes enter the rapid growth phase is unchanged. As we explain below, this mass-scale is determined by the ability of feedback from star formation to regulate the gas density within the galaxy, particularly in its centre, and is weakly dependent on the details of the black hole accretion physics.

\subsection{The buoyancy of star formation driven outflows}
\label{sec:buoyancy}

\begin{figure*}
\centering
\includegraphics[width=\linewidth]{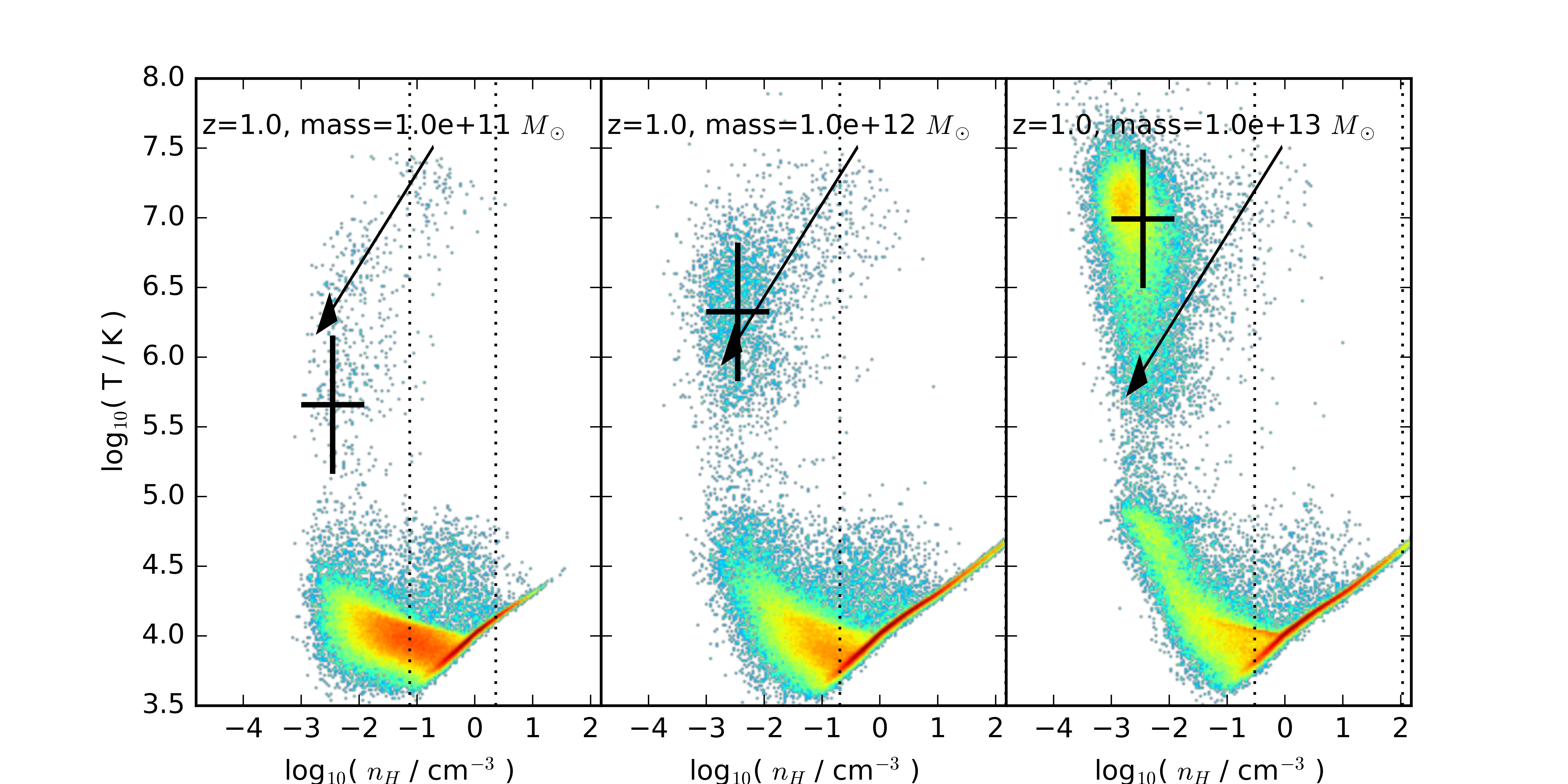}
\caption{This figure illustrates the buoyancy of the star formation-driven outflow  relative
to the galaxies' diffuse gas coronae. The three panels show how the temperature of the corona increases with the mass of the dark matter halo (from left to right). We show haloes at $z=1$; the situation is similar at other epochs. The background image illustrates the mass-weighted density--temperature distribution of the gas within the central  $0.1R_{\rm 200c}$ of haloes in the EAGLE simulation.  Crosses indicate the density and temperature of the hot corona used in the simple analytic model described in the text; the solid arrow shows the critical adiabat of the outflow (Eq.~\ref{eq:kwind}). In an outflow with order unity mass-loading, particles will be heated to the starting point of the arrow and then rise buoyantly,
expanding and adiabatically cooling along the arrow, until they reach pressure equilibrium. In $10^{11}\Msol$ haloes (left panel) the critical outflow adiabat (Eq.~\ref{eq:kwind}) comfortably exceeds that of the corona, allowing the outflow to be buoyant even if it has greater than unity mass loading.
In haloes of mass $10^{12}\Msol$ (central panel) and greater (right panel), the adiabat of the outflow is not buoyant and it is much more difficult for star formation to drive an efficient wind. 
The vertical dotted lines show the 10 and 90 percentiles of star forming gas density, confirming that the density of star forming gas increases with halo mass. 
}
\label{fig:halo_density_temperature}
\end{figure*}

The critical factor is therefore the ability of the star formation-driven outflow to effectively carry mass away from the galaxy, regulating its density and limiting black hole growth. Recent numerical simulations \citep{scannapieco2015} have shown that supernova cannot efficiently launch a ballistic wind and that most mass escapes the galaxy in a hot diffuse form. 
{\rgb 
In the absence of  a corona, gas hotter than the virial temperature will flow out as a rarefraction wave by converting its thermal energy into
a bulk flow.  This type of escape is seen in simulations of isolated disk patches \citep[eg.,][]{creasey2013}. The outflow leaves behind a diffuse 
corona which slowly cools. As the halo mass increases, and the characteristic cooling time of the halo becomes 
longer, the density of the corona increases \citep{white_frenk1991, birnboim2003}. Once the mass of the corona becomes comparable to that of the outflow, supernova heated gas must be buoyant compared to the surrounding gas corona in order to escape. (Although a sufficiently energetic outflow may sweep up and expel all of the halo material, this places more stringent requirements on the outflow.) 
}
If the corona is hot and dense (ie., formed of high entropy material), the outflow will not be buoyant and will stall, cool and fall back onto the galaxy. The buoyancy of the outflow is determined by the ratio of its `adiabat'\footnote{We use $k_B$ for the Boltzmann constant. The mean molecular weight of the plasma is
$\mu m_H$ where $m_H$ is the mass of hydrogen atom and $\mu \approx 0.59$ for a fully ionized plasma of primordial composition. Thermodynamic entropy is proportional the logarithm of the adiabat.}, $K = k_B T \left(\rho / \mu m_H\right)^{-2/3}$, to that of the galaxy's diffuse corona. 

To illustrate this, we begin by comparing the adiabat of the outflow to that of the halo. To determine the adiabat of the outflow, we assume that roughly half the energy available from supernovae is used to drive an outflow with a mass loading $\beta = \dot{M}_{\rm outflow}/\dot{M_*}$.
Specifically, we assigning the outflow a specific energy of $\sim 10^{49} \erg\Msol^{-1}$ (60\% of the supernova energy available from
a Chabrier initial stellar mass function, \citet[eg.,][]{crain2015}, and equivalent to a temperature of $3\times10^{7}\K$).
To estimate the density of the entrained/heated gas, we assume that galaxy disk sizes are set by the angular momentum of 
the dark matter halo \citep{mo_mao_white} so that $R_{\rm disk} \propto  (\Mhalo/10^{12}\Msol)^{1/3} \oneplusz^{-1}$  and that disk mass is proportional to the halo mass. (We use the notation $\oneplusz \equiv (\Omega_{\rm m} (1+z)^3 + \Omega_{\Lambda})^{1/3}$: at high redshift, where the effect of dark energy\footnote{$\Omega_{\rm m}$ and $\Omega_{\Lambda}$ are the cosmological density parameters for matter and dark energy respectively.}
is negligible, $\oneplusz \approx (1+z)$). The density then scales as $M_{\rm disk} R_{\rm disk}^{-2} \sim n_H^0 (\Mhalo/10^{12}\Msol)^{1/3} \oneplusz^{2}$ (where $n^0_H \sim 0.1\percm$ gives the average hydrogen atom density of ISM gas in present-day Milky Way-like galaxies) and the adiabat of the outflow is
\begin{equation}
K_{\rm outflow} \sim 8 \, \beta^{-1} \left(\frac{n^0_H}{0.1\percm}\right)^{-2/3} \left(\frac{\Mhalo}{10^{12}\Msol}\right)^{-2/9} \oneplusz^{-4/3}  \keVcm 
\label{eq:kwind}
\end{equation}
In order for the outflow to be a significant loss of mass from the galaxy disk we require that $\beta \ge 1$, and define a critical adiabat by setting $\beta=1$ so that $K_{\rm crit}\equiv  K_{\rm outflow}(\beta=1)$.

In contrast to the outflow, the adiabat of the galaxy's gas corona increases strongly with the mass of the halo:
assuming the inner parts of the corona have a temperature of
$T \approx 1.4\times10^{6} (\Mhalo/10^{12}\Msol)^{2/3} \oneplusz \,  {\rm K}$ in hydrostatic equilibrium, and a density of 30 times the times the baryonic virial density\footnote{We assume a density of 30 times $200 \rho_c \Omega_b$ where $\Omega_b$ is the baryon density parameter and $\rho_c$ is the critical density. The temperature we assume is twice the virial temperature of the halo
% ($T_{\rm vir} = 7.2\times10^{5}  \left(\Mhalo/1.0\time10^{12}\right)^{2/3} (1+z)$) 
since the radius of interest is close to the peak of the halo's rotation curve. We define the virial temperature as
$k_B T_{200}=\frac{1}{2}GM_{200}/R_{200}$, where $R_{200}$ is the radius enclosing an overdensity of 200 times $\rho_c$
and $M_{200}$ is the mass within it.}. 
we find
\begin{equation}
K_{\rm halo}\sim 7 \left(\frac{\Mhalo}{10^{12}\Msol}\right)^{2/3} \oneplusz^{-1} \keVcm.
\label{eq:khalo}
\end{equation}
In order for the corona to be in approximate hydrostatic equilibrium, $K_{\rm halo}$ must decline with radius; Eq.~\ref{eq:khalo}  provides a good description of the halo within 0.1 of the virial radius, as shown in  Fig.~\ref{fig:halo_density_temperature}.

Equating the critical outflow abiabat ($K_{\rm crit}$) with that of the halo (Eq.~\ref{eq:khalo}) picks out a critical halo mass scale of 
\begin{equation}  
M_{\rm crit} \sim  10^{12} \oneplusz^{-3/8} \Msol  
\label{eq:critical_mass} 
\end{equation}
In systems less massive than $M_{\rm crit} $, the outflow is buoyant, or escapes as a rarefraction wave, carrying away a significant fraction of the star forming gas, but in systems more massive than $M_{\rm crit} $, the outflow is either trapped by the surrounding corona (if $\beta>1$) or does not 
cause a significant loss of mass (if the outflow adjusts so that $K_{\rm outflow} > K_{\rm halo}$ this requires $\beta<1$). The gas that is not consumed by star formation then builds up in the central regions until rapid black hole growth is triggered. 
{\rgb
Our analysis does not consider the angular momentum carried away by the outflow: we assume that
tidal torques mediated by spiral density waves and clumping instabilities will carry gas into the centre of the system to feed the black hole.
}

\subsubsection{The halo mass dependence of the mass loading factor} 
The argument we have presented above explains the existence of a critical halo mass scale, above which the star formation is unable to drive an outflow with significant mass loading \citep[see also][]{keller2016}. In haloes less massive than $M_{\rm crit}$, the outflow will be buoyant if $\beta>1$, but we can take the argument a step further by assuming that $\beta$ adjusts 
so that the outflow is only just buoyant (ie., $K_{\rm outflow} \sim K_{\rm halo}$). This is plausible. When a parcel of gas starts to be heated it remains trapped until its adiabat is comparable to that of the halo. Once the adiabat exceeds $K_{\rm halo}$, however, 
the gas element will start to rise, so that it leaves the star forming region and is not heated further. As it leaves, it is replaced by another element, so that the energy output is spread over a greater gas mass. This is an overly simple interpretation of the complex physics of the interstellar medium, but it suggests that lower mass galaxies will drive outflows with a greater mass loading.

%
%
%% simple model for characteristic density
%
Using this argument, we can estimate the dependence of the mass loading on $M_{\rm halo}$. Equating this to the halo adiabat, we can estimate the halo mass dependence of the outflow:
\begin{equation}
\beta \sim \left(\frac{M_{\rm halo}}{M_{\rm crit}}\right)^{-8/9}. 
\label{eq:massloading}
\end{equation}
Note that the redshift dependence of the mass loading in this equation is encapsulated in the redshift dependence of $M_{\rm crit}$.

\subsubsection{A simple equilibrium model}
The mass scaling we have derived from the buoyancy argument above can be combined with a simple equilibrium model
to investigate the evolution of the host galaxy. The model we present is deliberately as simplified as possible and is presented
solely to motivate the choice of fitting functions used in the following section. A much more complete overview
of equilibrium models can be found in \cite{dekel_mandelker2014}.

The basis of equilibrium models is that the inflow of gas from the halo (roughly) balances star formation and outflow, ie.,
$
\dot{M}_{\rm in} = \dot{M}_*  + \dot{M}_{\rm out} .
$
Since we are concerned with low mass haloes, we assume that the mass inflow rate of gas is proportional to the halo growth rate, and that most of this mass flows directly onto the galaxy in cold streams. We also assume that the outflow is the dominant sink for this material
\begin{equation}
 f_b \dot{M}_{\rm halo} = (1-R+\beta) \dot{M}_* \sim \beta \dot{M}_*
\label{eq:equilibrium}
\end{equation}
where $f_b$ is the baryon fraction, and $R$ is the recycled stellar fraction.
We write the outflow mass loading as a function of halo mass,
$\beta = \left(M_{\rm halo}/ M_{\rm crit}\right)^{-\alpha}$, and note that power-law solutions of Eq.~\ref{eq:equilibrium} must be of the form:
\begin{equation}
M_* \propto \left(\frac{\Mhalo}{M_{\rm crit}}\right)^{1+\alpha}.
%%  M_* = f_b \left(\Mhalo \over M_{\rm crit}\right)^{1+\alpha}
\label{eq:mstar}
\end{equation}
Due to the logarithmic nature of the problem, these solutions are attractors, and if star formation and gas accretion are
sufficiently rapid, memory of the initial stellar mass is quickly forgotten.
Our analysis of the buoyancy of the outflow suggests $\alpha \approx 1$, and that the $M_*$ -- $\Mhalo$ relation should have a low mass slope close to $2$, as is indeed seen in observational data \citep[eg.,][]{moster2013,behroozi2013}. This lends credence to
our anzatz that the outflow will be only just buoyant.

\subsubsection{Relation to the central gas density}
\label{sec:bh_density_estimate}
In order to complete the analysis, we need to relate the outflow efficiency to the build-up of gas in the central regions of the galaxy.  We cannot expect to relate the global mass-loading factor $\beta$ directly to the density of gas on sub-kiloparsec scales around the black hole. Indeed, in \S\ref{sec:parameter_variations}, we show that a decrease in the normalisation factor $\kappa$ in the black hole accretion timescale is compensated by an increase in $\rho_{\rm bh}$. A full description of this self-regulation would be complex
(see \S\ref{sec:parameter_variations} for a discussion of the numerical simulations). 
 
We can, however, motivate a choice of scaling by assuming that the fraction of gas that is able to reach the central regions of the 
galaxy is inversely proportional to $\beta$. Using the gas density from Eq.~\ref{eq:kwind}, and rounding $8/9$ to $1$, this suggests
\begin{equation}
\begin{split}
n_{\rm bh} & =   n^0_{\rm bh} \left(\frac{M_{\rm halo}}{M_{\rm crit}}\right)^{8/9} \left(\frac{\Mhalo}{10^{12}\Msol}\right)^{1/3} \oneplusz^2\\
              & \sim  n^0_{\rm bh} \left( \frac{M_{\rm halo}}{10^{12}\Msol}\right)^{4/3}  \oneplusz^{5/2}.
\end{split}
\label{eq:rhoBH}
\end{equation}
where $n_{\rm bh}$ is the gas density around the blackhole, expressed in Hydrogen atoms per cm$^3$, and
$n^0_{\rm bh}$ is a normalisation constant.

We need to turn to the reference EAGLE simulation to see if the scaling suggested above is reasonable in practice, and to determine the normalisation constant. In \S\ref{sec:bh_density}, Fig.~\ref{fig:density_redshift_mass} shows that the density surrounding the black hole is indeed reasonably well described by Eq.~\ref{eq:rhoBH}. In the simulation, the slope of the mass dependence 
appears to increase at with redshift.  This can be qualitatively understood if gas collects rapidly at in the
central regions once the outflow becomes in efficient. The effect of adopting a redshift dependent slope is to trigger rapid
growth in slightly lower mass haloes at high redshift, making BH trajectories (in the $\Mbh$ -- $\Mhalo$ plane) more similar.

Fig.~\ref{fig:density_redshift_mass} shows that a suitable choice of normalisation is $ n^0_{\rm bh} = 3 \cm^{-3}$.
With this choice of normalisation, black holes in present-day $10^{12}\Msol$ haloes have growth time, $t_{\infty} = 11.3 \Gyr$.

\subsection{Black hole growth in high mass haloes} 
\label{sec:bh_mass_limit}

The focus of this paper is to explain the rapid growth of black holes at the transition mass scale, $M_{\rm crit}$, and their low accretion rates below it. Above this mass scale, we assume that the mass of the black hole is limited to a fraction of the halo binding energy. Using the total halo binding energy, however, does not take into account the long cooling times of massive haloes in the present-day Universe. We therefore compute the binding energy within
a halo mass dependent radius.  For the relevant halo masses, the cooling rates of metal enriched plasmas are not particularly sensitive to temperature, so that the cooling time will roughly scale as 
\begin{equation}
t_{\rm cool} \propto (1+z)^{-3} \left( \frac{r}{ R_{200}} \right)^2 T_{200},
\end{equation}
where $R_{200}$ and $T_{200}$ are the halo virial radius and temperature respectively.  Equating the cooling time
to the age of the Universe at redshift $z$ suggests that the radius within which we should calculate the binding
energy scales as roughly $\Mhalo^{-1/3} (1+z)^{1/4}$ relative to the virial radius. We then calculate the binding energy within this radius (allowing for the variation in concentration with halo mass \citep{duffy2008}) to compute the 
upper limit to the black hole mass. Clearly this estimate is approximate, and our approach is deliberately simple. A more advanced analytic treatment would account for low entropy gas expelled from the system \citep{mccarthy2011} and the impact of the subsequent rearrangement on the halo cooling time. The treatment would also need to account for the growth of black holes due to black hole -- black hole mergers. All of these effects are, however,  accounted for in the EAGLE numerical simulations.

\subsection{Linking black hole growth and halo growth}
\label{sec:halo_growth}

Today's supermassive black holes are thought to originate from black hole `seeds' of mass
$10^4$ to a few times $10^5\Msol$ that are created from the direct collapse of metal free gas clouds \citep[eg.,][]{begelman2006,regan2009, mayer2010}.  We translate the growth of black holes into a relation between black hole mass and halo mass by assuming that black holes are created as seeds in haloes capable of hosting a small galaxy, and then comparing the black hole growth rate with that of its host dark matter halo. 
In a $\Lambda$-CDM cosmology, the halo growth rate is well approximated by \citet{correa2015}
\begin{equation}
\dot{M}_{\rm halo} = 7 \times 10^{10}  \left(\frac{\Mhalo}{10^{12}\Msol}\right)  (0.51+0.75 z) \oneplusz^{3/2} \Msol\Gyr^{-1}. 
\label{eq:halogrowthrate}
\end{equation}

%%galaxy wind will stall at a particular halo mass
The equilibrium model allows us to link the halo growth rate, the effectiveness of star formation driven outflows and the gas density in the galaxy. 
The density of gas surrounding the black hole depends critically on the effectiveness of this feedback from star formation. If star formation-driven outflows are not effective at limiting the density of the ISM, the black hole will be driven into its rapid growth phase. A 100-fold increase in gas density translates into an order of magnitude decrease in the timescale for black hole growth, and the timescale $t_{\infty}$ would then fall from being longer than the age of the Universe to being a small fraction of it. Black hole growth and the effectiveness of star formation-driven outflows are thus intertwined.

%% predicted tracks in Mhalo-Mbh
Combining Eqs.~\ref{eq:halogrowthrate} and~\ref{eq:rhoBH} enables us to plot the trajectory of a black hole in the black hole mass -- halo mass plane. To match the EAGLE simulations, we assume that black holes have an initial seed mass of $1.4\times 10^{5}\Msol$.
Seeding black holes at different times creates a variety of trajectories, which we illustrate by coloured lines in Fig.~\ref{fig:bh_growth_model}. The transition halo mass at which this rapid growth phase occurs is rather weakly dependent on the seeding time, and the spread in trajectories is further compressed if the trajectory is plotted as a function of stellar mass. %
%% predicted relation at the present day.
By connecting the final black hole masses, we predict the black hole mass -- halo mass relation at the present day (shown as a black dashed line). The model is idealised, and we would expect scatter in both the growth rates of haloes and the density of gas surrounding the black hole. This will lead to the populations of red and blue galaxies overlapping at the transition mass scale: red systems having significantly more massive black holes than blue galaxies of equal halo mass (as seen in Fig.~\ref{fig:sfgrowth_mstar}).

%% observable predictions of the model - galaxy transition mass scale doe snot depend on redshift
The model we have presented makes some other remarkable observable predictions. Firstly, it suggests that the transition mass scale should be weakly dependent on redshift. This is indeed confirmed by observations that measure the fraction of red sequence galaxies as a function of mass (as can be seen by comparing Fig.~\ref{fig:sfgrowth_mstar} with the equivalent plot for present-day galaxies in \citet{kauffmann2003}). 
%% prediction the bh-mass, galaxy mass relation will be non-linear
Secondly, the model predicts that the relation between black hole mass and galaxy mass will be strongly non-linear, rising steeply below a galaxy mass of $\sim 3\times 10^{10}\Msol$, before flattening to an almost linear relation. 

\begin{figure*}
\centering
\includegraphics[width=\linewidth]{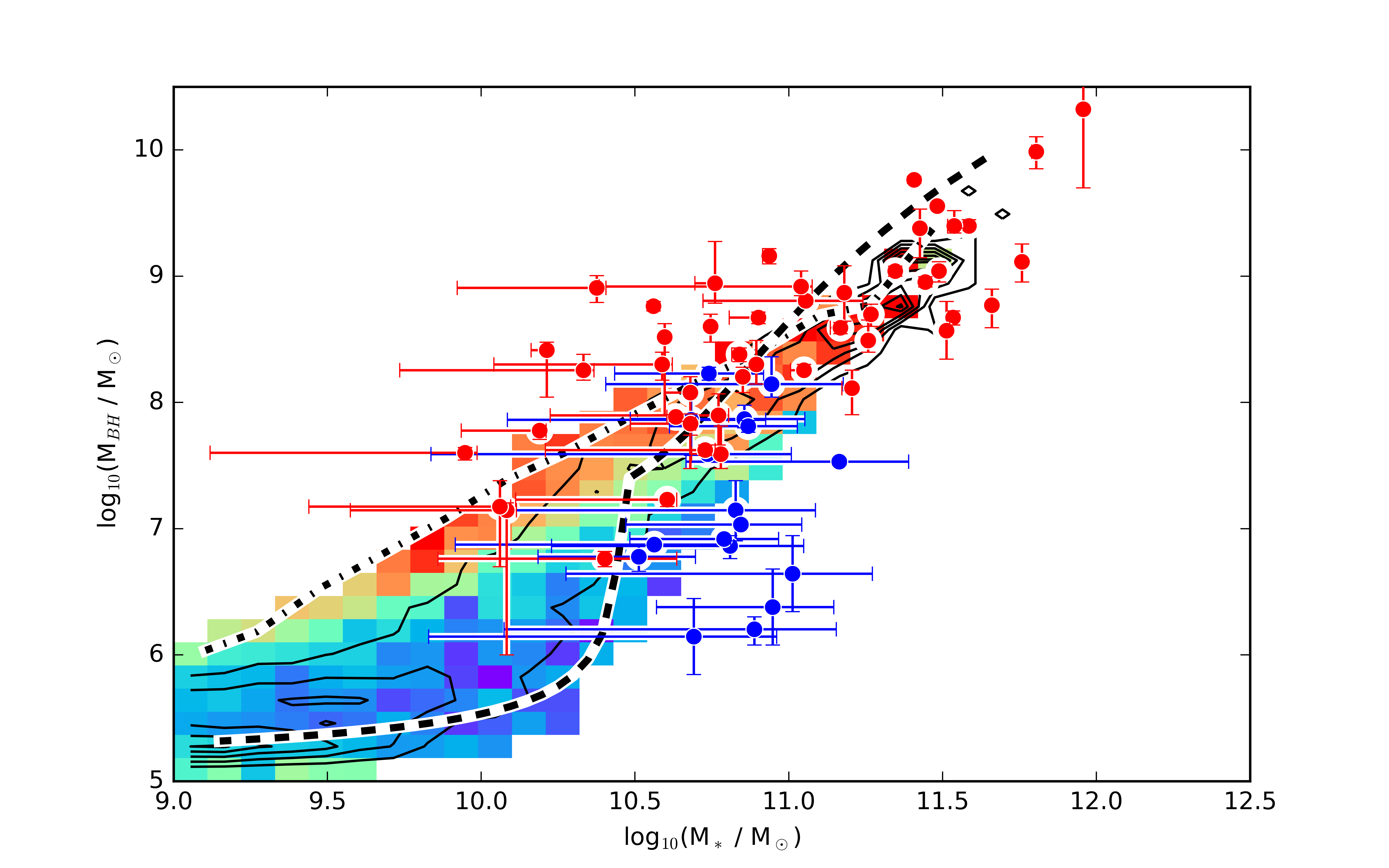}
\caption{The observed and predicted relation between black hole mass and total galaxy stellar mass. The relation predicted by the simple analytic model is shown as a black dashed line. This is obtained from the equivalent line in Fig.~\ref{fig:bh_growth_model} using the observationally inferred relation between galaxy mass and mass of the dark matter halo mass \citep{moster2013}. 
%Thinner coloured lines illustrate the evolution of two example black holes. 
Dynamical measurements of black hole masses in nearby galaxies are shown as coloured points with error bars \citep{savorgnan2016}, with red and blue points indicating early and late-type galaxies respectively. Vertical errors indicate uncertainties in the black hole mass, while horizontal error bars extend from the galaxy bulge mass to the effect of assuming a maximal disk stellar mass. 
The background image shows galaxies from the EAGLE simulation, with the colour showing the average star formation growth timescale in each pixel and contours summarising the galaxy number density in the $\Mbh$ -- $M_*$ plane. The hydrodynamic simulation follows the behaviour of both the analytic model and the observations, validating the assumptions we have made. Furthermore, the dot-dashed line shows the median relation 
we obtain when we re-run the simulation turning off feedback from star formation. In the absence of star formation driven outflows, black holes are always able to accrete efficiently
and grow along a power-law relation.
}
\label{fig:model}
\end{figure*}

We compare the prediction of the analytic model to observational data in Fig.~\ref{fig:model}, using the observationally derived relation between halo mass and galaxy mass \citep{moster2013} to convert the black hole mass -- halo mass relation predicted by the analytic model into a relation between black hole mass and galaxy mass. The prediction of the analytic model is shown as a thick dashed line. The observational data are shown as red and blue points, depending on the
host galaxy morphology (red and blue corresponding to early and late-type morphologies, respectively). A key point in this plot is that
recent observational developments have made it possible to measure the masses of black holes over a wide range of galaxy mass and morphology \citep[eg.,][]{mcconnell2013, savorgnan2016}. To plot total galaxy mass, rather than than bulge mass alone, we convert the mid-IR photometry given by \citet{savorgnan2016} to stellar mass using the calibration of \citet{meidt2014}. The error bars on the point show the uncertainty in this conversion. The lower limit shows the bulge mass, while the upper limit assumes there is no non-stellar contribution to the emission. The data points correspond to the assumption that 50\% of the disk emission is from diffuse gas in late-type systems, and 10\% in early-type systems. When plotted in this way, the non-linear relation between black hole mass and galaxy mass becomes clear, as does the connection between black hole mass and galaxy morphology. The observations confirm the model's prediction that early-type (red sequence) galaxies harbour more massive black holes than late-type (blue sequence) galaxies of equivalent mass.

Clearly the analytic model makes the prediction that isolated galaxies with masses below the current limit of the observation data will have black hole masses less than $10^6\Msol$. Unfortunately, it is extremely hard to directly measure the masses of such small black holes. Further more, it is critical to correctly account for non-detections and to avoid contamination of the sample by tidally stripped satellite galaxies \citep{barber2015}. By estimating black hole masses from the emission line widths, \citet{reines2015} give us a glimpse of the relation at lower masses. The results are fully consistent with the predictions of the analytic model.

\section{Validating the Analytic Model with the EAGLE Hydrodynamic Simulations}
\label{sec:validation}

\subsection{The Eagle simulation suite}
\label{sec:eagle_simulations}
In order to validate our simple analytic model, we have analysed the growth of black holes in the state-of-the-art EAGLE cosmological simulations \citep{schaye2015,crain2015}. The simulation suite consists of a large number of hydrodynamical
simulations of galaxy formation, including different resolutions, simulated volumes and physical models. The simulations
use advanced smoothed particle hydrodynamics (ANARCHY, \citet{schaller2015}), based on the Gadget code \citep{springel2005},  and phenomenological subgrid models to capture unresolved physics including cooling, metal enrichment and energy input from stellar feedback and black hole growth \citep{dallavecchia2008,wiersma2009a,wiersma2009b}. A complete description of the code can be found in \citet{schaye2015}.  The simulations successfully reproduce key observational properties of galaxies, including the evolution of the galaxy stellar mass function, galaxy sizes and the transition mass scale for galaxy star formation rates \citep{schaye2015,furlong2015,furlong2016,trayford2016}. The simulations also provide a reasonable description of the observed black hole X-ray luminosity function \citep{rosas2016}.
The simulations uses $\Lambda$-CDM cosmological parameters taken from the Planck Collaboration 2014 measurements \citep{planck2014}, specifically $\Omega_{\rm m}=0.307$, $\Omega_\Lambda=0.693$, $\Omega_b=0.04825$, $h=0.6777$ and $\sigma_8= 0.8288$.

\begin{table*}
\begin{center}
\renewcommand{\arraystretch}{1.2}
\begin{tabular}{lrrrrrrrrrrrr}
\hline
Identifier & L & N & $m_{\rm{g}}$ & $\gamma_{\mathrm{eos}}$ & $n_{\rm{n}}$ & $C_{\rm{visc}}$ & $\Delta \rm{T}_{\rm{AGN}}$ & $m_{\mathrm{seed}}$\\
& [cMpc] & & [M$_{\odot}$] & & & [K] & [M$_{\odot}$] \\
\hline\hline
REFERENCE & 100 & $2{\times}1504^{3}$ & $1.81{\times}10^{6}$ & 4/3 &  2/ln10& 2$\pi$ & $10^{8.5}$ & $1.475 \times 10^{5}$ \\
EagleVariation\_AGNdT9p00\_ALPHA1p0e4 & 50 & $2{\times}752^{3}$ & $1.81{\times}10^{6}$ & 4/3 &  2/ln10& {$\bf 2\pi{\times} 10^2$} & $10^{8.5}$ & $1.475 \times 10^{5}$ \\
EagleVariation\_SEED1p0e4 & 50 & $2{\times}752^{3}$ & $1.81{\times}10^{6}$ & 4/3 &  2/ln10& 2$\pi$ & $10^{8.5}$ & {$\bf 1.475 \times 10^{4}$} \\ 
EagleVariation\_ALPHA1p0e4 & 50 & $2{\times}752^{3}$ & $1.81{\times}10^{6}$ & 4/3 &  2/ln10& {$\bf 2\pi{\times} 10^2$} & $10^{8.5}$ & $1.475 \times 10^{5}$ \\
EagleVariation\_ALPHA1p0e8 & 50 & $2{\times}752^{3}$ & $1.81{\times}10^{6}$ & 4/3 &  2/ln10& {$\bf 2\pi{\times} 10^{-2}$} & $10^{8.5}$ & $1.475 \times 10^{5}$ \\
EagleVariation\_ONLY\_AGN & 50 & $2{\times}752^{3}$ & $1.81{\times}10^{6}$ & 4/3 & {\bf --} & 2$\pi$ & {$\bf 10^{8.5}$} & $1.475 \times 10^{5}$ \\
EagleVariation\_EOS1p666& 25 & $2{\times}376^{3}$ & $1.81{\times}10^{6}$ & {\bf 5/3} &  2/ln10& {$\bf 2\pi{\times} 10^2$} & $10^{8.5}$ & $1.475 \times 10^{5}$ \\
HIRES-REFERENCE & 25 & $2{\times}752^{3}$ & $2.26{\times}10^{5}$ & 4/3 &  2/ln10& 2$\pi$ & $10^{8.5}$ & $1.475 \times 10^{5}$ \\
HIRES-RECALIBRATED & 25 & $2{\times}752^{3}$ & $2.26{\times}10^{5}$  & 4/3 &  {\bf 1/ln10}& {$\bf 2\pi{\times} 10^3$} & {$\bf 10^{9.0}$} & $1.475 \times 10^{5}$ \\
\hline
\end{tabular}
\end{center}
\caption{Parameters describing the available simulations. From
  left-to-right the columns show: simulation name; comoving box size;
  total number of particles; initial baryonic particle mass 
  %dark matter particle mass; comoving Plummer-equivalent gravitational softening length;  maximum physical softening length 
  and the subgrid model parameters that vary: $\gamma_{\mathrm{eos}}$ 
  %$n_{\rm{H,0}}$, 
  $n_{\rm{n}}$, $C_{\rm{visc}}$ and $\Delta
  \rm{T}_{\rm{AGN}}$ (see section 4 of \citet{schaye2015} for an explanation of their
  meaning). The changes are highlighted in bold. Simulations with initial baryonic mass $1.81{\times}10^{6} \Msol$ ($2.26{\times}10^{5}\Msol$) have dark matter particle mass $9.70{\times}10^{6} \Msol$ ($1.21{\times}10^{6} \Msol$).}
\label{tab:sim_table}
\end{table*}

The largest EAGLE simulation, which we label REFERENCE and use in most plots, uses a simulation carried out in a comoving volume of $(100\Mpc)^3$ using $1504^3$ baryon particles of initial mass $1.81\times10^6\Msol$ and the same number of dark matter particles of mass $9.70\times10^6\Msol$.  The simulation uses a gravitational softening that is no larger than $0.70\pkpc$ (we use the notation $\pkpc$ to clearly distinguish proper length scales from comoving length scales). A convergence study is shown in Appendix~A: we find that the results we present are not changed by a factor of 8 decrease in the particle mass. We use additional simulation models to explore the impact of different subgrid physics assumptions. These are listed in Table~\ref{tab:sim_table}. The simulation outputs were analysed using the SUBFIND programme to identify bound sub-structures \citep{springel2001,dolag2009} within each dark matter halo.
We identify these substructures as galaxies and measure stellar masses within a radius of $30 \pkpc$. The black hole
mass we quote corresponds to the central black hole of each galaxy.

\subsection{Black holes and galaxies in the EAGLE simulations}
\label{sec:blackholes_in_eagle}

%% computer simulations - existence of early-late- dichotomy and relation to BH mass
In the EAGLE simulations, the accretion on to a black hole is determined 
by a sub-grid model that accounts for the mean density, effective sound speed and relative motion and angular momentum of the surrounding gas as detailed in \cite{rosas2015} (with the exception that, in the EAGLE simulations, we do not increase the accretion rate to account for an unresolved clumping factor). The model assumes that once gas reaches high densities on sub-kpc scales around the black hole, it will be accreted on to the black hole at the Bondi rate unless its angular momentum is sufficiently high to prevent this. Our simulations allow us to understand how the density on sub-kpc scales around the black hole is 
determined by the interaction of star formation, feedback and gas accretion (on scales of 1~kpc and greater) without making the simplifications adopted in the analytic model. Importantly, the simulation does {\it not} impose a galaxy transition mass scale by varying the black hole feedback efficiency as a function of halo mass or accretion rate. We always assume a that the energy generated by 
feedback is 1.5\% of the rest mass energy of the accreted material.

The reference EAGLE simulation accurately predicts the bimodality of galaxy types and the existence of a tight and rapidly evolving sequence of star formation growth timescales in blue-sequence galaxies \citep{furlong2015,trayford2016}. This is illustrated in Fig.~\ref{fig:sfgrowth_mstar}, where coloured points show a selection of galaxies from the simulation. The simulation also reproduces many aspects of the observed AGN population well, including the creation of a population of X-ray bright AGN in the first billion years of the Universe \citep{rosas2016}.

In Fig.~\ref{fig:model}, we show galaxies sampled from the simulation as contour lines and a background image coloured by the average star formation growth timescale, $\dot{M}_*/M_*$. The simulation follows both the analytic model and the observations well, reproducing the red-/blue-sequence dichotomy. Furthermore, the colouring of the image shows that the transition is driven by black hole growth: at a given mass, galaxies lying on the blue (star forming) sequence have black holes an order of magnitude smaller than equivalent red-sequence (passive) galaxies, showing the importance of black hole growth in this transition region.
{\rgb
The agreement on the transition mass scale between the analytic model and the simulation is encouraging since the choice of the parameter $n^0_{\rm bh}$ was based on the gas density around black holes (in $10^{12}\Msol$ haloes) in the simulation, and not fit to reproduce the transition mass seen in the simulation. 
%% numerical experiments
The greatest value of the hydrodynamic simulations is, however, that we can vary the parameters that control feedback from star formation and the accretion rates of black holes, and thus experiment with their impact on the resulting galaxy and black hole correlations.
In the following section, we will explore how the value of $n^0_{\rm bh}$ changes in response to variations in the simulation parameters, gaining deeper insight into the connection between the galaxies corona and the density of gas around the black hole. 
}

\subsection{Density of gas around the black hole}
\label{sec:bh_density}

\begin{figure}
\centering
\includegraphics[width=1.1\columnwidth]{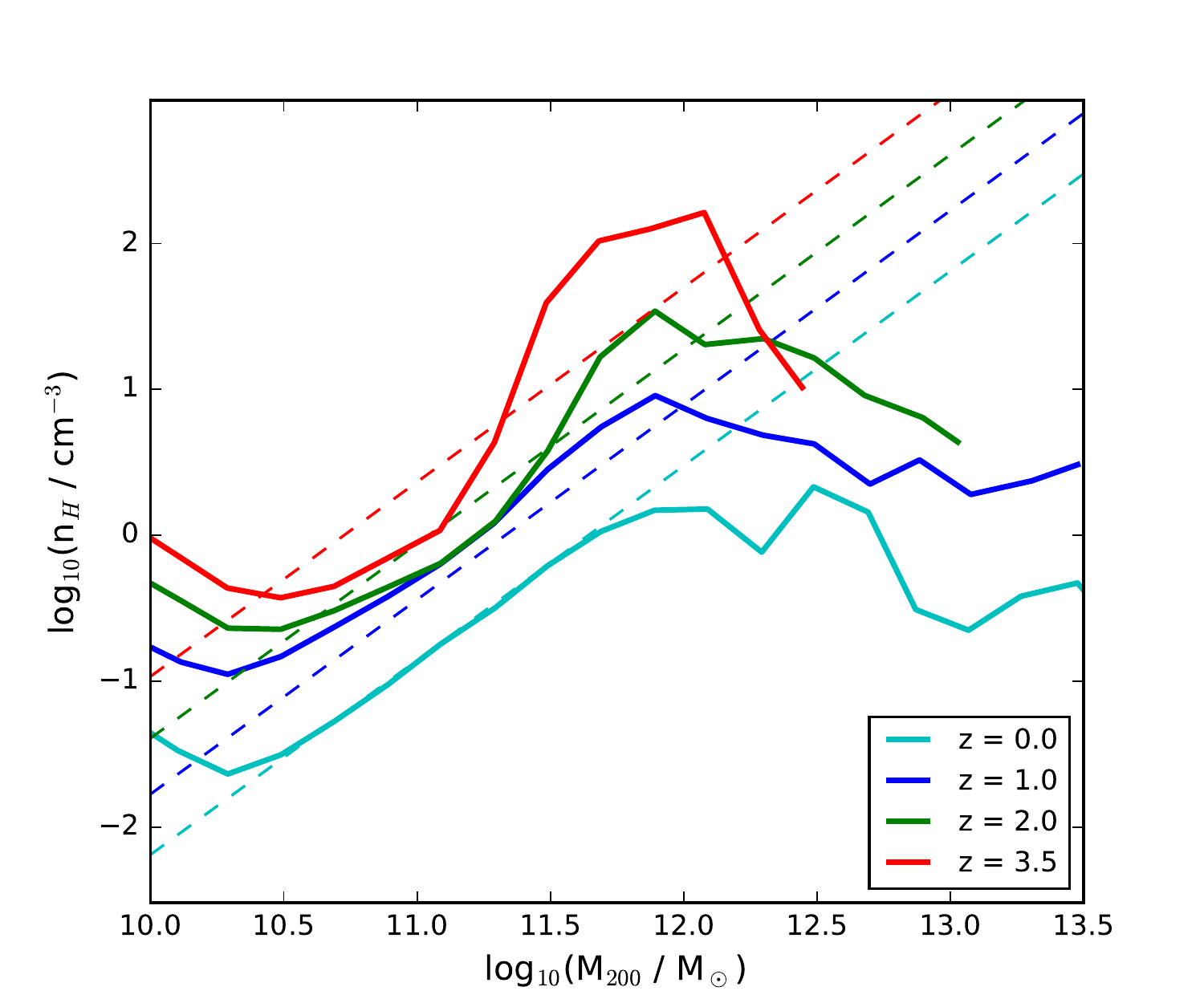}
\caption{
The figure shows the average density of gas around the black hole (on $\sim 700\pc$ scales) as a function of halo mass in the EAGLE simulation. Individual black holes show significant scatter around this relation. The different line colours show how the relation evolves, with higher densities being apparent at higher redshifts. The density of the gas increases rapidly between $10^{11}$ and  $10^{12}\Msol$, as we would expect if it is driven by the confinement of outflows. The rise results in a rapid increase in accretion rate. The resulting AGN feedback has a profound impact on the properties of the host galaxy, as shown in Fig.~\ref{fig:sfgrowth_mstar}.  Dashed lines in the plot illustrate the simple analytic model discussed in \S\ref{sec:bh_density_estimate}. The simulation follows the main features of the analytic description, but displays a
more complex behaviour, with the slope of the mass dependence evolving with redshift. Above a halo mass of $10^{12}\Msol$, the density declines as the energy 
input from the black hole deprives the galaxy of gas.
}
\label{fig:density_redshift_mass}
\end{figure}

The ability of the galaxy to launch an outflow is part of a complex story that cannot be completely captured in a simple analytic model, and we  turn to the full hydrodynamic simulation to support the functional form adopted for the density of gas around the black hole on $\sim 700\pc$ scales (the resolution of the reference simulation). Fig.~\ref{fig:density_redshift_mass} shows 
that the density of material surrounding the black hole is a strong function of both redshift and halo mass. 
The high densities reached at high redshift bring the average black hole accretion rates close to the Eddington limit, and allows the model to explain the presence of high redshift quasars \citep{mortlock2011,rosas2016}.  

The functional form suggested by the analytic model (dashed lines) matches that in the simulation reasonably well, capturing the main features, particularly at lower redshift. At higher redshift, the simple analytic model underestimates the build-up of the gas density in haloes more massive than $M_{\rm crit}$ seen in the simulation. This has the effect of reducing the spread in the halo mass at which black holes enter their rapid growth phase, leading to a  situation in which the most rapidly growing black holes are found in haloes of mass $\sim 10^{12}\Msol$, regardless of redshift.  
Black holes in lower-mass haloes have yet to enter the non-linear growth phase, while black holes in higher-mass haloes have injected so much energy into the host halo, re-arranging its density and entropy structure \citep[eg.,][]{mccarthy2011}, that inflowing gas is unable to remain cool and provide further fuel for star formation \citep[eg.,][]{dubois2013}.

\subsection{Parameter variations}
\label{sec:parameter_variations}

\begin{figure*}
\centering
\hspace*{-3.cm}
\includegraphics[width=0.5\linewidth]{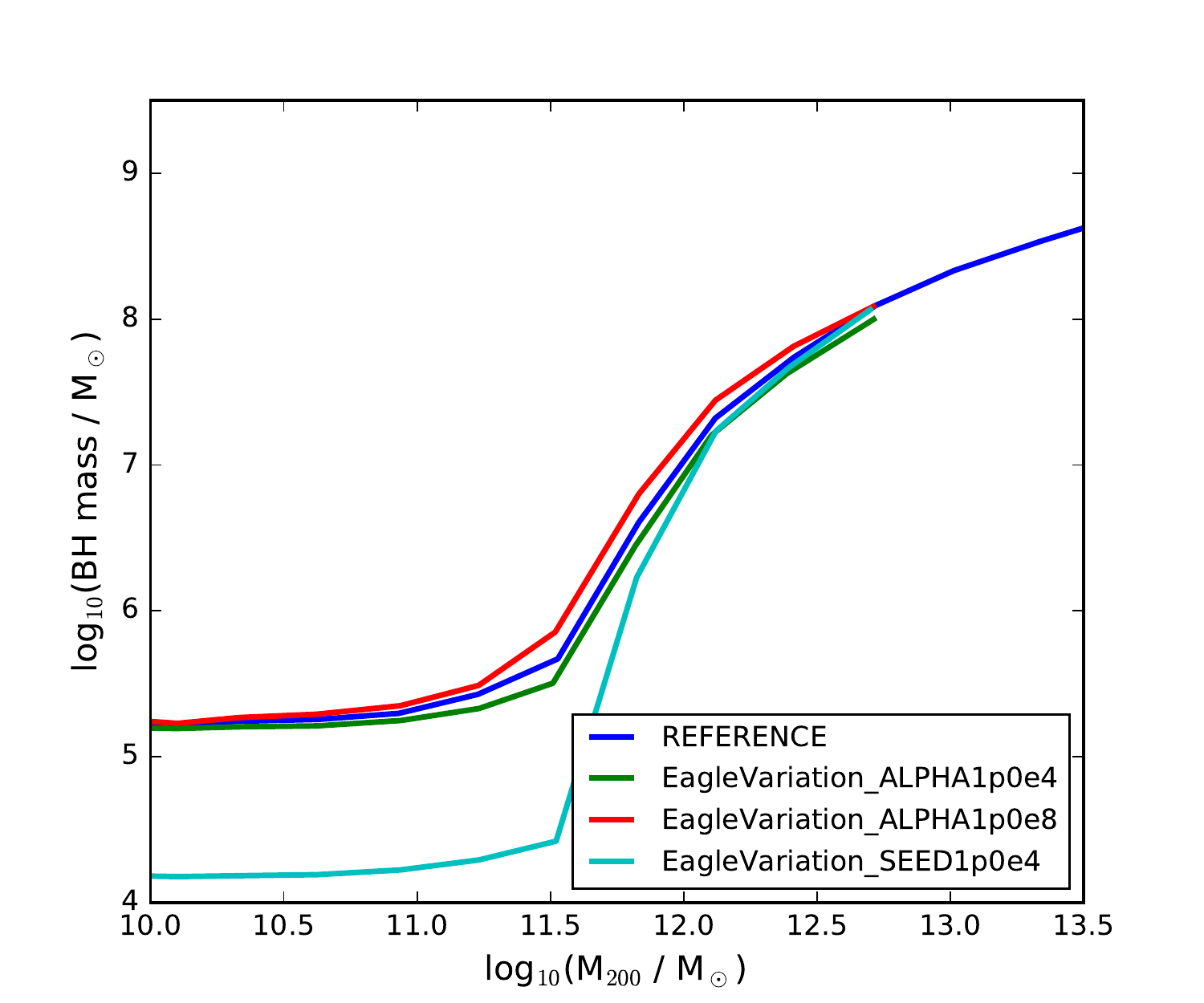}
\hspace*{-0.1cm}
\includegraphics[width=0.5\linewidth]{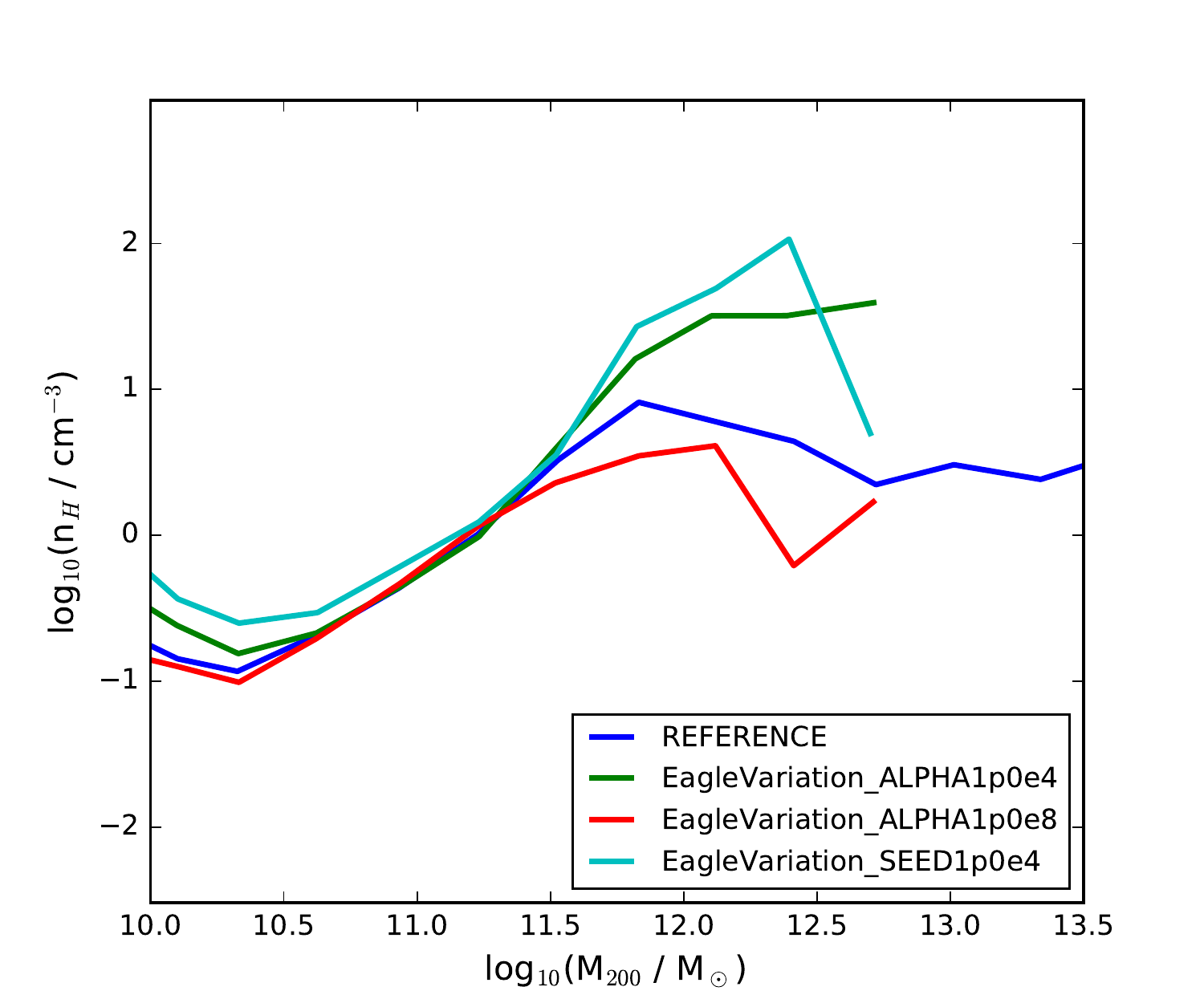}
\hspace*{-3.cm}
\caption{
The {\it left-hand panel} shows relation between central black hole mass and the mass of the halo at $z=1$. The coloured lines show models in which the AGN accretion parameter and black hole seed mass are varied. The existence of a transition mass scale is remarkably insensitive to these parameters, with all the models showing a rapid transition at the same mass scale. The {\it right-hand panel} shows the average density of gas around the central black hole (on $\sim 700\pc$ length scales) as a function of halo mass. The line colours correspond to the models in the left panel. The difference between the lines provides an explanation for the similarity of the black hole masses in the models: variations in the normalisation of the accretion rate are compensated by an increase or decrease in the peak density of gas around the black hole. 
}
\label{fig:bh_model_variations}
\end{figure*}
The EAGLE simulation suite includes many variations of model parameters. In the model we have presented, the accretion rate onto the black hole undergoes a transition that is driven by the hot halo of the galaxy. This predicts that the transition mass scale should be insensitive to the star formation law and the details of the AGN accretion and feedback model. In Fig.~\ref{fig:bh_model_variations}, we show three models in which the AGN accretion parameter and black hole seed mass are varied. These models are discussed in more detail in \citet{crain2015} and \citet{salcido2016}, and were run in smaller, $(50 \Mpc)^3$, volumes compared to the REFERENCE simulation (but with the same resolution). 
In models EagleVariation\_ALPHA1p0e8 and  EagleVariation\_ALPHA1p0e4, the $\alpha$ parameter controlling the angular momentum dependence of the black hole accretion rate (and thus the coefficient $\kappa$ in Eq.~\ref{eq:t_infinty}) is increased or decreased by a factor 100, respectively. In the left hand panel, we show the relation between halo mass and the central black hole mass. Despite varying the normalisation of the black hole accretion rate parameter by 4 orders of magnitude, the models show remarkable similarity to the reference EAGLE simulation. The initial black hole masses in the EagleVariation\_SEED1p0e4 simulation are a factor 10 smaller than in the reference run. Despite this, massive black holes are able to grow, following a similar relation to that seen in the REFERENCE simulation.  The robustness of the model is remarkable, confirming our analytic model in which the growth of the black hole, the steep transition in the black hole mass-halo mass relation, and the transition in galaxy star formation rates are driven by the properties of the halo, and are not a coincidence of the details of the black hole accretion parameters. 

The right-hand panel of Fig.~\ref{fig:bh_model_variations} provides an explanation of this behaviour, showing the density of the gas surrounding the black hole as a function of halo mass in the same models. Models in which the black hole accretion rate normalisation is lower, or the seed black hole mass is smaller, reach higher densities, compensating for the lower normalisation of the initial accretion rate. This is the behaviour expected in our model: when the star formation-driven outflow ceases to be effective in higher-mass haloes, the gas density in the centre of the galaxy builds up until efficient black hole accretion and feedback occur. 

\begin{figure*}
\centering
\hspace*{-3.cm}
\includegraphics[width=0.5\linewidth]{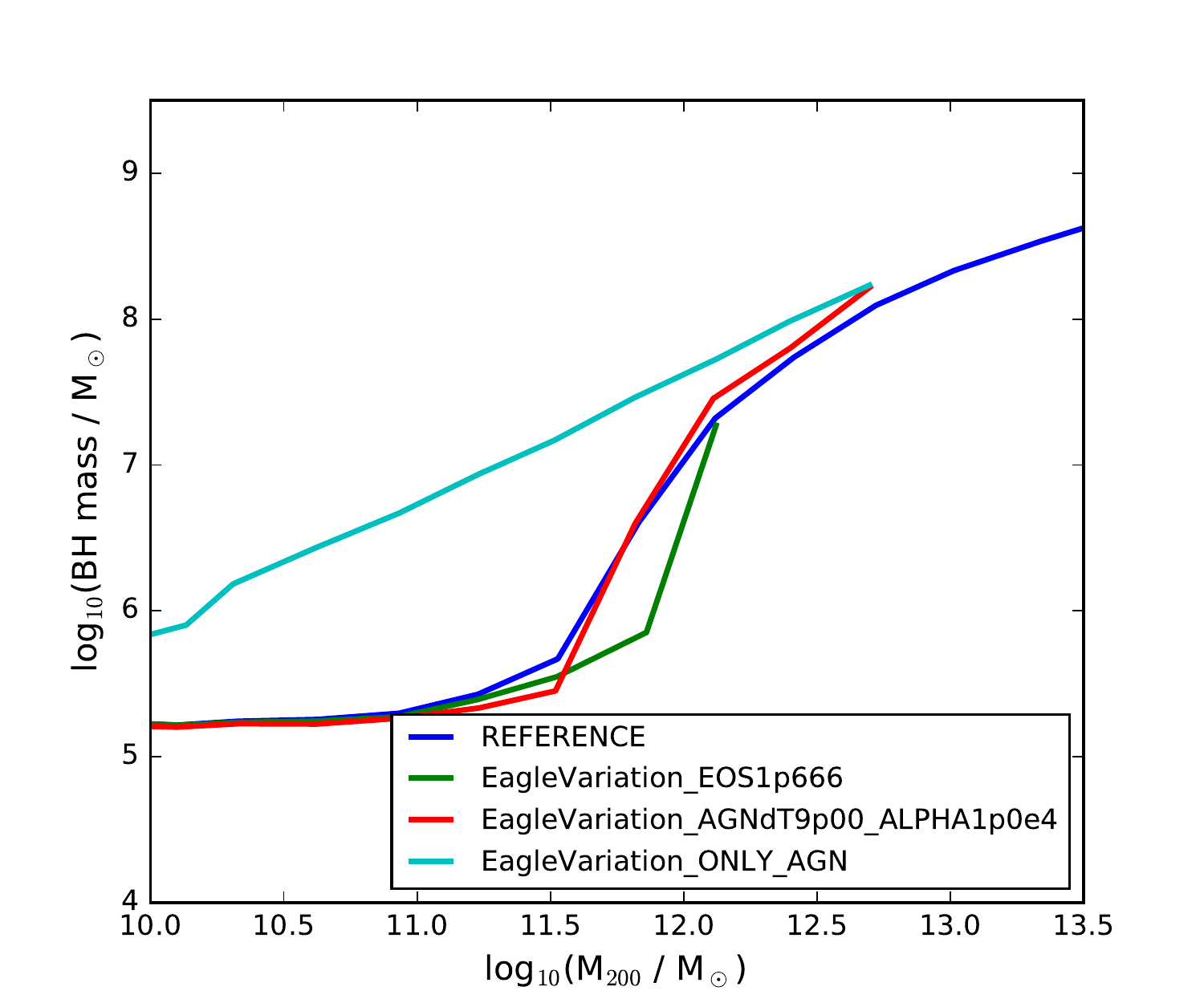}
\hspace*{-0.1cm}
\includegraphics[width=0.5\linewidth]{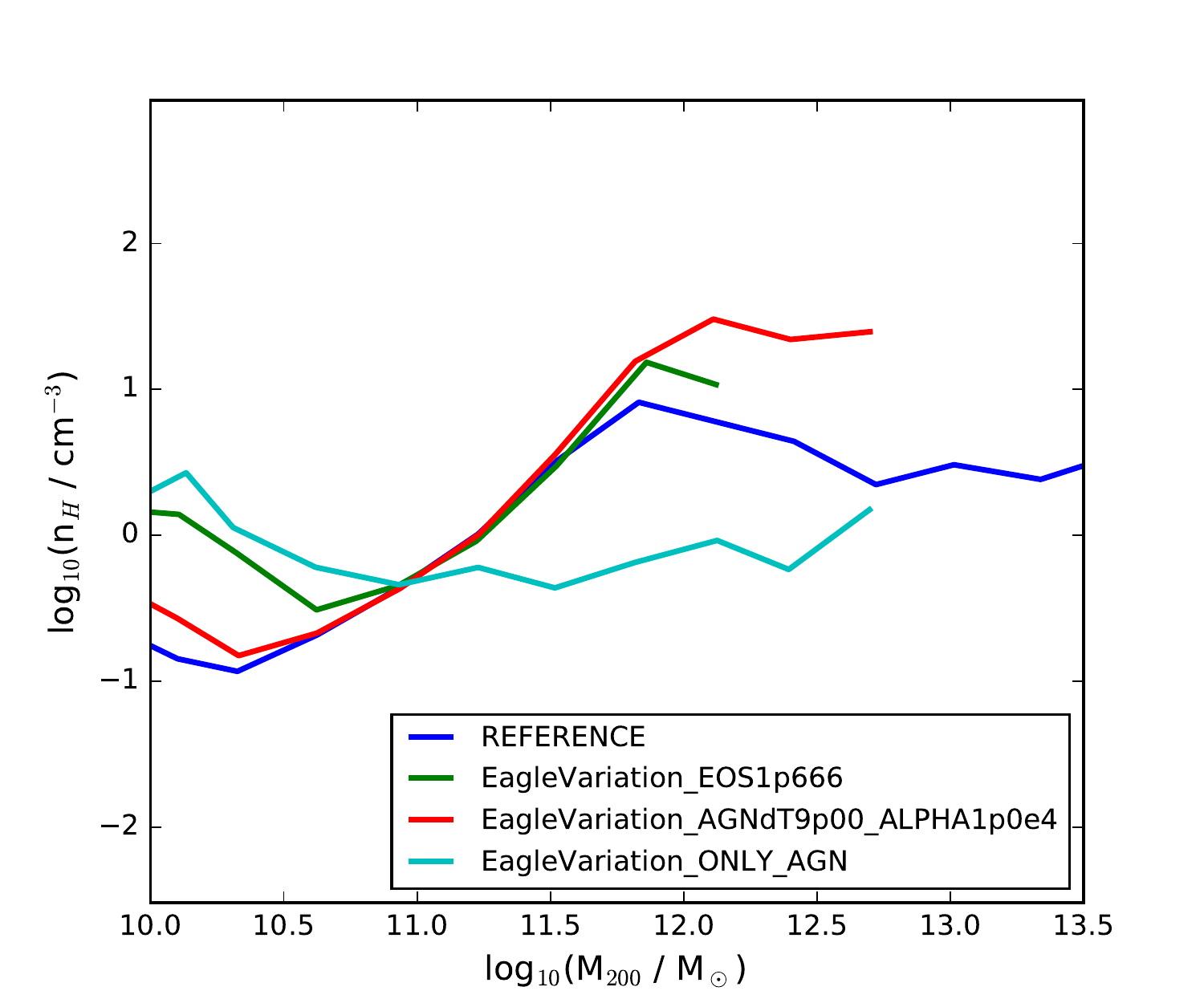}
\hspace*{-3.cm}
\caption{
The panels in this figure are the same as in Fig.~\ref{fig:bh_model_variations}, but here we compare the reference simulation
with simulations that vary the AGN feedback heating temperature and the ISM effective equation of state, as well as a model 
in which only AGN feedback is included (so that no star formation-driven outflow is present). The black hole mass -- halo mass relations in the first two variations are remarkably similar, and the right hand panel shows that densities around the black holes follow similar trends. This is expected since neither the detailed modelling of the ISM, nor the implementation of the AGN feedback, affect the halo gas adiabat which determines whether the star formation driven outflow can regulate the central gas density. In contrast, the model in which energetic feedback from star formation is not implemented shows completely different behaviour. In particular, there is no break in the black hole mass -- halo mass relation and black holes grow effectively in low-mass haloes so that their mass is limited by the halo binding energy rather than by the 
gas supply. This behaviour is predicted by the analytic model we have presented.}
\label{fig:more_model_variations}
\end{figure*}

In Fig.~\ref{fig:more_model_variations}, we show the central black hole mass and black hole gas density relations for three further model variations. The simulations EagleVariation\_EOS1.666 and EagleVariation\_AGNdT9\_ALPHA1p0e4 show the effects of varying the exponent of the equation of state assumed for the ISM and the heating temperature used in the AGN feedback model. These simulations were run in $(25 \Mpc)^3$ and $(50 \Mpc)^3$ boxes, respectively, using the same resolution as the REFERENCE simulation. The increase in the equation of state exponent, from $4/3$ to $5/3$, reduces the compressibility of the ISM. Despite its impact on the internal properties of galaxies, the assumed equation of state has relatively little impact on the transition mass scale since the gas reaches similar densities around the black hole. The higher AGN heating temperature in simulation EagleVariation\_AGNdT9\_ALPHA1p0e4 was introduced in order to increase the abiabat of gas heated by AGN feedback (but not its overall energy), but this also has little impact on the masses of black holes or the density of the gas that builds up around them. This model does, however, result in more realistic X-ray properties in galaxy groups \citep{schaye2015}. 

Fig.~\ref{fig:more_model_variations} shows a third model, in which stellar feedback is switched off. This model is also shown as a dot-dashed line in Fig.~\ref{fig:model}.   The resulting black hole masses and densities are markedly different in this case. Since stellar feedback is no longer removing gas from small galaxies, the density around the black hole builds up regardless of the presence or absence of a hot, high-entropy corona. As a result, no transition in the masses or activity of the black hole is seen. The model does {\it not} match the observed stellar (or black hole) properties of galaxies, clearly demonstrating the need for
a transition between stellar and AGN feedback in setting the properties of galaxies and establishing the existence of the characteristic mass scale in galaxy formation.

\section{Discussion and Conclusions}
\label{sec:discussion}

Galaxy properties show an abrupt change in behaviour around a halo mass of $10^{12}\Msol$. In lower-mass haloes, central galaxies are rapidly star forming, doubling their stellar mass through star formation on a timescale that is shorter than the age of the Universe. In higher-mass haloes, galaxies have much longer star formation growth timescales (Fig.~\ref{fig:sfgrowth_mstar}) and grow primarily through galaxy mergers and the accretion of external stars. This dichotomy is often referred to as the blue and red galaxy sequences, respectively. The change in star formation rate at the transition mass scale leads to a change in slope of the galaxy mass -- halo mass relation, creating a break in the galaxy mass function at a galaxy mass scale of $3\times10^{10}\Msol$. The decline in the importance of on-going star formation and the rise in the importance of growth by mergers may also account for the shift from late to early-type galaxy morphology at this mass scale. Explaining the origin of this mass scale is of fundamental importance to explaining the structure of the observable Universe.

In this paper, we have developed a simple analytic model in which 
black holes and star formation compete to regulate the gas content of a galaxy as it grows by accretion from the cosmic web. The main elements of the model are the non-linear growth of black holes in the Bondi accretion regime and the buoyancy of star formation - driven outflow relative to 
galaxies hot coronae.  

In the Bondi regime, black holes grow slowly before abruptly switching to a rapid growth phase (Fig.~\ref{fig:bh_growth_model}). The onset of this phase
is highly dependent on the surrounding gas density, and thus on the ability of star formation driven outflows to prevent gas accumulating in the centre of the galaxy. In our model, the effectiveness of stellar feedback depends on the buoyancy of 
the outflow relative to the corona (in the absence of a corona, the outflow escapes as a rarefraction wave). We show that this leads to a critical halo mass above which star formation
driven outflows are unable to prevent the build up of gas. Making simple assumptions about the evolution of the gas density of star forming galaxies, we compare the buoyancy of the star formation driven outflow to that of the halo (Fig.~\ref{fig:halo_density_temperature}) in order to determine
the dependence of the central gas density on halo mass and redshift. The model links the build up of the gas density in the central regions of the galaxy, the on-set of rapid black hole growth, the galaxy's eventual transition to the red sequence and the build-up
of the hot corona.

In order to test the simplified analytic model, we compare the model to the growth of black holes in the EAGLE hydrodynamic simulations. The EAGLE simulations assume a constant efficiency of black hole feedback, but at the same time obtain a remarkable match to the observed properties of galaxies, including the abrupt change in galaxy star formation rates in the transition regime (Fig.~\ref{fig:sfgrowth_mstar}).
Although no halo mass dependence of black hole accretion rates is imposed in the simulations, the transition in galaxy properties emerges as we would expect from the analytic model (Fig.~\ref{fig:model}). 

The greatest value of the hydrodynamic simulations is, however, that we can vary the parameters that control feedback from star formation and the accretion rates of black holes, and thus experiment with their impact on the resulting galaxy and black hole correlations. This enables us to confirm the causal connections implied by the simple analytic model.  For a fixed density, varying the implicit accretion disk viscosity, or the seed black hole mass, alters $\kappa$, and hence the 
onset of the rapid growth phase, $t_{\infty}$. In practice, however, reductions in $\kappa$ are compensated by an increasingly steep relation between the gas density around the black hole and halo mass, and the galaxy transition mass is remarkably insensitive to the parameter
choice (Fig.~\ref{fig:bh_model_variations}). Variations in the effective equation of state assumed for star-forming gas or the heating temperature of AGN feedback (Fig.~\ref{fig:more_model_variations}) also have little impact on the transition mass scale. If we reduce or eliminate feedback from supernovae, however, we find that black holes are able to grow effectively in haloes of all masses, as illustrated by the dot-dashed line in Fig. ~\ref{fig:model} and the cyan line in Fig.~\ref{fig:more_model_variations}. This change eliminates the transition mass scale and consequently does not reproduce the properties of observed galaxies. This confirms that, 
below the transition mass, black hole growth is suppressed by stellar feedback.

%% summary 
Galaxies fall into two distinct galaxy types, characterised by the presence or absence of significant on-going star formation. This dichotomy is established in the early universe and driven by a transition in galaxy properties at a mass of $\sim 3\times 10^{10}\Msol$. We have presented a new way of understanding this transition and, thus, the origin of the distinct red and blue galaxy sequences. Rapid growth of black holes is triggered at a halo mass of $\sim 10^{12}\Msol$ by the development of a sufficiently hot diffuse gas corona which confines the star formation-driven outflow by preventing it from rising buoyantly.  This simple analytic model is supported by observational data and confirmed in numerical experiments. It makes predictions for the relations between the growth of galaxies and their black holes which will be tested by forthcoming observations. In particular, the
model predicts that the black holes present in isolated low-mass galaxies will be small, with most objects having black holes smaller than $10^6\Msol$. 

\section*{Acknowledgements}

We gratefully acknowledge Oliver Ilbert and COSMOS team for assistance with the star formation rate data for young galaxies,
and Alistair Graham and Sandra Savorgnan for assistance with the stellar and black holes masses of local galaxies.
This work was supported by STFC grant ST/L00075X/1 and used the DiRAC Data Centric system
at Durham University, operated by the Institute for Computational Cosmology on behalf of the STFC DiRAC HPC
Facility (www.dirac.ac.uk). This equipment was funded by
BIS National E-infrastructure capital grant ST/K00042X/1,
STFC capital grant ST/H008519/1, and STFC DiRAC
Operations grant ST/K003267/1 and Durham University.
DiRAC is part of the National E-Infrastructure. We also
gratefully acknowledge PRACE for awarding us access to
the resource Curie based in France at Trs Grand Centre de Calcul. This work was sponsored by the Dutch National Computing Facilities Foundation (NCF) for the use
of supercomputer facilities, with financial support from the
Netherlands Organization for Scientific Research (NWO).
The research was supported in part by the European Research Council under the European Union's Seventh Framework Programme (FP7/2007-2013) / ERC Grant agreements 278594-GasAroundGalaxies, GA 267291 Cosmiway,
and 321334 dustygal, the Interuniversity Attraction Poles
Programme initiated by the Belgian Science Policy OWNce
([AP P7/08 CHARM]), the National Science Foundation under Grant No. NSF PHY11-25915, the UK Science and Technology Facilities Council (grant numbers ST/F001166/1 and
ST/I000976/1), Rolling and Consolidated Grants to the
ICC, Marie Curie Reintegration Grant PERG06-GA-2009-256573,
% Marie Curie Initial Training Network Cosmocomp (PITN-GA-2009-238356). 
RAC is a Royal Society University Research Fellow. RGB, JS and RAC thank the Simons Foundation for their hospitality at the `Galaxy Superwinds' symposium where this work was completed. The data used in this project is available from the EAGLE public data base at http://icc.dur.ac.uk/Eagle/database.php \citep{mcalpine2016}.

\bibliographystyle{mnras}
\bibliography{dark_nemisis_of_galaxy_formation_mnras}

\appendix

\section{Numerical Convergence}

\begin{figure*}
\centering
\hspace*{-3.cm}
\includegraphics[width=0.5\linewidth]{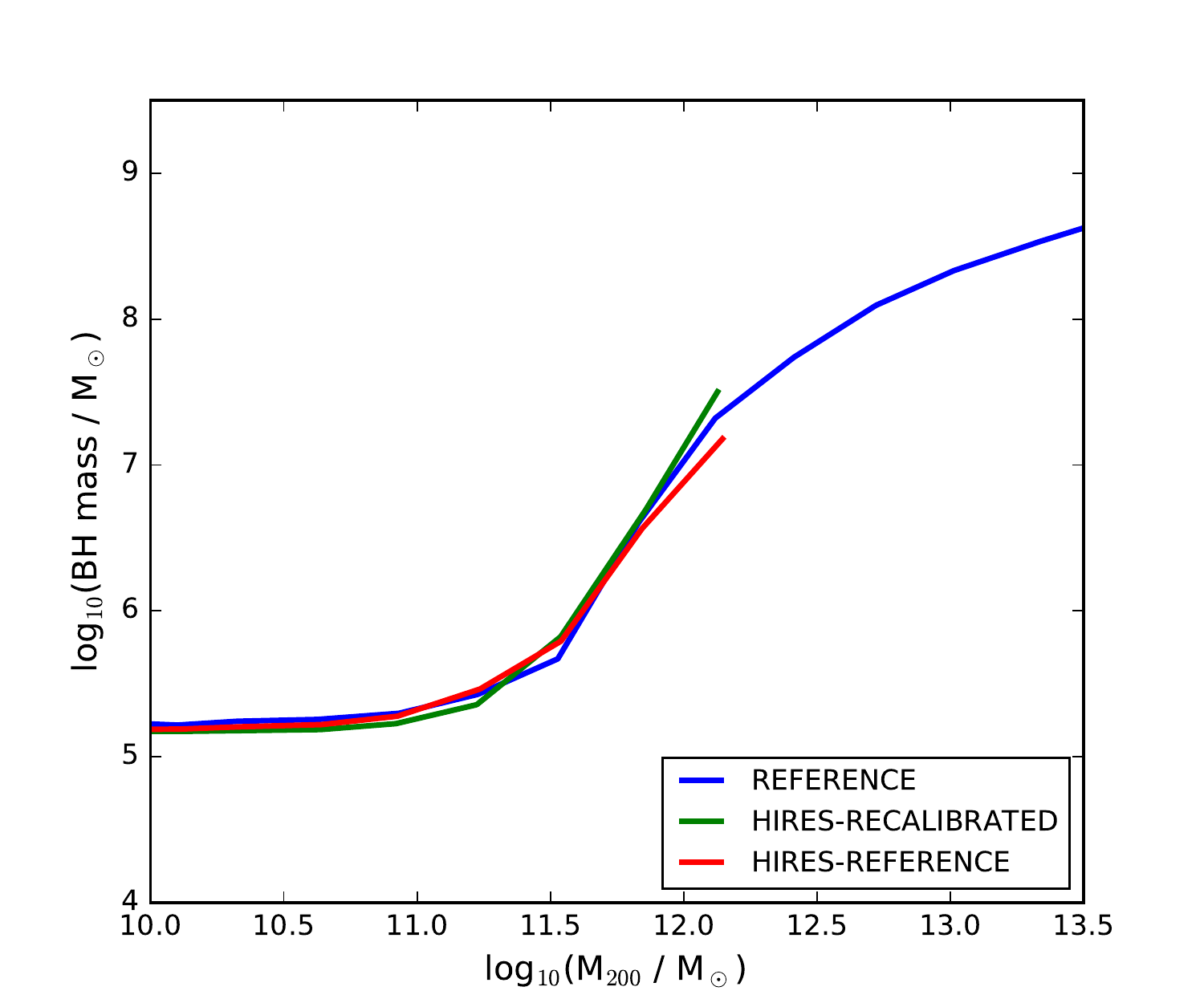}
\hspace*{-0.1cm}
\includegraphics[width=0.5\linewidth]{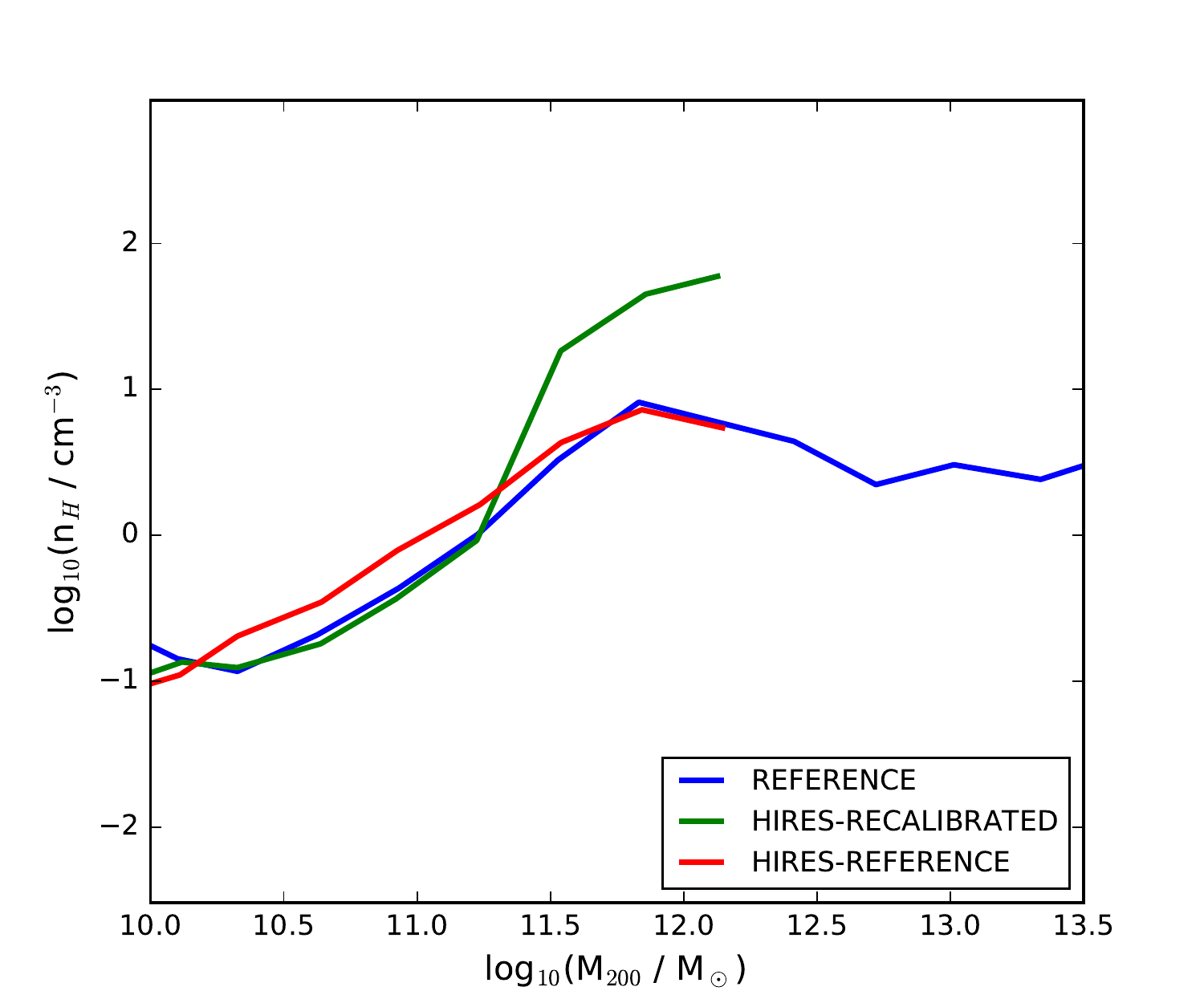}
\hspace*{-3.cm}
\caption{
The panels in this figure are the same as in Fig.~\ref{fig:bh_model_variations}, but in this figure we compare the reference simulation
with two higher resolution versions of the simulation allowing us to investigate the weak and strong convergence of the numerical methods. Both versions of the high resolution simulations yield black hole mass -- halo mass relations that are very close to the REFERENCE simulation in the left-hand panel. The differences between the high resolution simulations seen in the right-hand panel
correspond to the differences expected from the parameter variation study in \S~\ref{sec:parameter_variations} of the supplementary information.}
\label{fig:simulation_convergence}
\end{figure*}

For the reference EAGLE simulation, we are able to investigate the effect of repeating simulations at higher numerical resolution.  We compare REFERENCE to two simulations in which the particle mass is 8 times smaller (and the gravitational softening a factor of 2 smaller) in Fig.~\ref{fig:simulation_convergence},
reproducing the black hole mass and black hole density panels discussed for the model variations in \S\ref{sec:parameter_variations}. The higher resolution simulations were carried out in a $(25 \Mpc)^3$ volume. Using the nomenculture introduced in \citet{schaye2015}, the simulations test the `strong' convergence (in which we keep all parameters fixed at their fiducial values) in HIRES-REFERENCE and the `weak' convergence of the simulation in  HIRES-RECALIBRATED. As discussed by Schaye et al., a calculation with higher resolution resolves additional physical processes and that may require recalibration of the parameters encapsulating the subgrid processes. In the left hand panel, which shows the black hole mass -- halo mass relation, both approaches to convergence yield similar results, and agree extremely well with the REFERENCE calculation on which our parameter investigations are based. In the right hand panel, which shows the gas density around the black hole as a function of halo mass, the two higher resolution simulations differ. This is expected since the $\alpha$ parameter, controlling the angular momentum dependence of the accretion rate, is a factor of 100 smaller in HIRES-RECALIBRATED than in the reference run. This confirms that the parameter dependencies investigated at the resolution of the REFERENCE simulation also hold in the higher-resolution simulations.

\end{document}